%% file: main.tex
\pdfoutput=1  % tell arXiv's AutoTeX to use pdflatex -- must stay in first 5 lines

\documentclass[11pt]{article}

\usepackage{styles/arxiv}
\usepackage{styles/remora}

\draftfalse   % draft notes OFF for the final build
\usepackage[T1]{fontenc}
\usepackage[utf8]{inputenc}
\IfFileExists{lmodern.sty}{%
  \usepackage{lmodern}%
  \usepackage{microtype}%
}{%
  \IfFileExists{mathptmx.sty}{\usepackage{mathptmx}}{}%
  \usepackage[expansion=false]{microtype}%
}

\usepackage{amsmath,amssymb,amsfonts}
\usepackage{booktabs}
\usepackage{tabularx}
\usepackage{colortbl}
\usepackage{multirow}
\usepackage{makecell}
\usepackage{graphicx}
\usepackage{float}
\usepackage[font=small,labelfont=bf,skip=6pt]{caption}
\usepackage{subcaption}

\usepackage{algorithm}
\usepackage{algpseudocode}

\usepackage{enumitem}
\usepackage{url}
\usepackage[hidelinks,breaklinks=true]{hyperref}
\IfFileExists{xurl.sty}{\usepackage{xurl}}{}
\usepackage{cleveref}
\usepackage{orcidlink}

\graphicspath{{figures/}}

\usepackage[numbers,sort&compress]{natbib}
\title{REMORA: A Voice-First Memory Assistant for Grounded\\ Personal Recall and Independent Living}

\shorttitle{REMORA: A Voice-First Memory Assistant}
\renewcommand{\headeright}{Preprint}

\author[1]{Bilal Jawaid\thanks{Corresponding author: \texttt{bilaljawaidcs@gmail.com}}}
\author[1]{Muhammad Maaz Uddin}
\author[1]{Osama Farooqi}
\author[1]{Saadia Karim}
\affil[1]{Department of Computer Science, Salim Habib University, Karachi, Pakistan}

\date{}

\begin{document}
\maketitle

\input{sections/00_abstract}

%%%%%%%%%%%%%%%%%%%%%%%%%%%%%%%%%%%%%%%%%%%%%%%%%% body %%%%%%%%%%%%%%%%%%%%%%%

\input{sections/01_introduction}
\input{sections/02_related_work}
\input{sections/03_design_goals}
\input{sections/04_system_design}
\input{sections/05_retrieval_generation}
\input{sections/06_implementation}
\input{sections/07_evaluation_method}
\input{sections/08_results}
\input{sections/09_discussion}
\input{sections/10_threats_to_validity}
\input{sections/11_limitations_future_work}
\input{sections/12_conclusion}
\input{sections/13_backmatter}

%%%%%%%%%%%%%%%%%%%%%%%%%%%%%%%%%%%%%%%%%%%%%%% references %%%%%%%%%%%%%%%%%%%%

\bibliography{refs/references}

%%%%%%%%%%%%%%%%%%%%%%%%%%%%%%%%%%%%%%%%%%%%%%% appendices %%%%%%%%%%%%%%%%%%%%

\appendix
\input{appendix/A_survey_instrument}

\input{appendix/B_pilot_questionnaire}

\input{appendix/C_pilot_scenarios}

\input{appendix/D_prompt_templates}

\input{appendix/E_reproducibility}
\input{appendix/F_interface}

\end{document}

%% file: sections/00_abstract.tex
\begin{abstract}
Most of what matters in a day is spoken and written down nowhere: a dosing
interval named at a clinic, a deadline moved in passing, the place a document was
set down. Such information is consequential because it is operational, and
fragile because it arrives during divided attention---the moment its importance
becomes clear is rarely the moment it was said. For people living with mild
cognitive impairment, Alzheimer's disease or age-related episodic memory decline,
that gap bounds independent living. Conventional aids close it poorly: notes,
reminders and calendars require the user to notice, interrupt and transcribe,
adding extraneous cognitive load exactly where attention is already committed.
We present \sysname, a voice-first memory assistant that converts consented
speech into structured, auditable memory records. \sysname transcribes and
diarizes captured audio, extracts memory candidates---tasks, appointments, names,
instructions, object locations---each retaining provenance into the
conversation it came from, and holds records, embeddings and relationships in a single
relational system of record. Recall combines lexical, semantic, relational and
recency signals, since no one signal covers the ways people interrogate their own
past, and reaches the user either through an explicit query or through a
confidence-gated cue that stays silent unless it is likely to help. Generation is
bound by a grounding contract: only retrieved evidence may be used, and absent a
supporting memory \sysname abstains rather than confabulates, because a fluent
invention about one's own past will be acted upon as fact. A requirements survey
and a pilot study with memory-loss users indicate that conversational forgetting
extends well beyond clinical populations, that grounded and concise recall is
feasible in controlled settings, and that privacy, rather than accuracy, is the
binding constraint on adoption.
\end{abstract}

\keywords{memory augmentation \and assistive technology \and cognitive load \and
episodic memory \and mild cognitive impairment \and retrieval-augmented generation
\and hybrid retrieval \and speaker diarization \and grounded generation \and
abstention \and human--AI interaction \and independent living}

%% file: sections/01_introduction.tex
\section{Introduction}
\label{sec:introduction}

\begin{figure}[htbp]
  \centering
  \includegraphics[width=\textwidth]{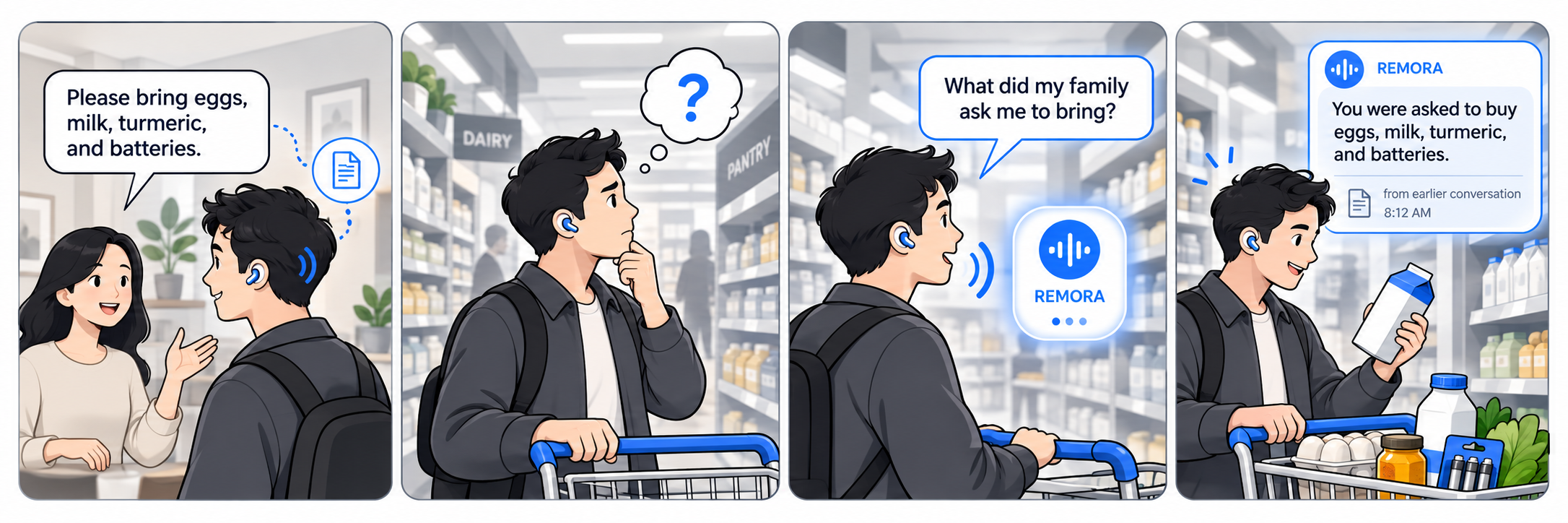}
  \caption{A representative recall episode. \textbf{(1)}~A request is made in
    conversation and captured with the speaker's consent. \textbf{(2)}~The detail
    is needed hours later, at the moment it has been forgotten. \textbf{(3)}~The
    user asks for it in a single short question rather than searching for a note.
    \textbf{(4)}~\sysname{} answers from the stored record and returns the
    conversation and time the answer came from, so the user can check it against
    the moment it was spoken.}
  \label{fig:scenario}
\end{figure}

Much of the information that governs a person's day is spoken once and recorded
nowhere: a physician names a dosing interval on the way out of the room, a
colleague assigns a task in the closing seconds of a meeting, a family member
lists what to buy on the way home. Such utterances are
\emph{operational}---they determine what someone must do next---and
\emph{ephemeral}: they arrive during divided attention and leave no artefact
behind. By the time such information matters, the only surviving copy is in
memory. \Cref{fig:scenario} traces one such episode from the moment it is spoken
to the moment it is needed.

For most people, losing that copy is an inconvenience; for a large and growing
population it constrains independent living. Dementia affected an estimated 57
million people worldwide in 2019 and is forecast to reach 153 million by
2050~\citep{gbd2022dementia}, with care for those already affected the immediate
obligation~\citep{livingston2024lancet}. Mild cognitive impairment---decline that
does not yet compromise functional independence~\citep{petersen2011mci}---most
often presents in an amnestic form in which episodic memory is affected first,
and longitudinal neuroimaging links structural brain change to that
decline~\citep{vidalpineiro2025vulnerability}. The deficit is asymmetric in a way
that matters for design: retention of recent conversational detail degrades
earliest, while the ability to \emph{act on} that detail once supplied is
comparatively preserved. That asymmetry is the premise of this work. \sysname is
assistive, not therapeutic, and makes no claim to diagnose or treat disease.

Conventional aids are misaligned with the moment of need. Notebooks, reminder
applications and voice memos presuppose that the user will \emph{notice} that
something matters, \emph{interrupt} what they are doing, and \emph{transcribe} it
accurately---three demands on the attention the conversation is already
consuming. Cognitive load theory predicts the
outcome~\citep{sweller2024cognitive}: the aid adds extraneous load to a task
already near working-memory capacity and is abandoned precisely when it would
help. The first step is also the one impairment attacks first, since it requires
remembering that something is worth remembering.

Capturing speech automatically removes that burden, but imposes three
requirements that must hold together. The system must be \textbf{trustworthy}:
generative models produce fluent statements that no source
entails~\citep{huang2025hallucination}, and here the user holds no independent
copy against which to check them, so retrieving evidence before
answering~\citep{lewis2020rag} must be joined by a willingness to report that the
answer is unknown. It must be \textbf{deployable} on the phones and browsers
people already own, rather than on hardware the intended users cannot readily
obtain or tolerate. And it must be \textbf{acceptable}: what it captures belongs
partly to other people, making its appropriate use context-dependent in a way
current models handle poorly~\citep{mireshghallah2024confaide}.

\sysname (Reliable Enhanced Memory-Oriented Recall Assistant) is built around
these three requirements. It captures consented speech through ordinary web and
mobile clients, transcribes and diarizes it, extracts memory
candidates---tasks, appointments, names, instructions, object locations---and
stores each as a structured record carrying provenance back to its source
transcript segment. Retrieval combines lexical, semantic, relational and recency
signals over a single relational system of record; generation answers only from
retrieved evidence and abstains when none supports the question.

Read against \Cref{fig:scenario}, the four panels correspond to the four
commitments the rest of this paper defends. Capture is consented and ambient, so
nothing is demanded of the user at the moment of speaking (panel~1). Recall is
initiated by a short question rather than a search, so the cost of remembering
stays below the cost of the lapse (panel~3); where the need can be inferred from
context, a confidence-gated cue may surface it without being asked, and stays
silent otherwise. The answer is returned with the conversation and time it came
from (panel~4): that timestamp is not decoration but the provenance link that
lets the user check the system against their own past. And when no stored record
supports the question, the same mechanism produces a refusal instead of a
plausible sentence.

We contribute a memory architecture in which every record carries provenance to
its source segment, held in one relational store rather than split across
relational, vector and graph systems; a hybrid retrieval formulation for personal
memory, together with the query classes on which each signal alone fails; an
interaction model that specifies when assisted recall stays silent; and a
grounding contract treating abstention as a measured outcome, evaluated through a
requirements survey and a pilot study with memory-loss participants.

%% file: sections/02_related_work.tex
\section{Related Work}
\label{sec:related}

\sysname{} sits at the intersection of four lines of work: memory augmentation
and lifelogging, LLM-based memory interfaces, memory architectures for
autonomous agents, and the grounding literature that governs whether a retrieved
answer can be trusted. We take each in turn and state where \sysname{} departs.

\subsection{Memory Augmentation and Lifelogging}
\label{sec:related:lifelog}

Externalising memory through capture has a long lineage in ubiquitous computing.
SenseCam established that passively captured images, reviewed afterwards, act as
retrieval cues strong enough to support recall in people with memory
impairment~\citep{hodges2006sensecam}, and later multimedia-cue systems extended
the idea to richer signals and to everyday rather than clinical
use~\citep{dingler2016multimedia}. This work established the premise \sysname{}
inherits: a durable external trace, created without effort at the time of
encoding, can substitute for a failed internal one.

It also exposed the costs. Continuous visual capture produces material that must
be browsed rather than queried, imposes a hardware burden, and raises objections
from bystanders who did not choose to be recorded. \sysname{} narrows the
capture channel from imagery to consented speech and replaces post-hoc browsing
with direct question answering, so that the retrieval effort the lifelogging
tradition pushed onto review time is absorbed by the system.

\subsection{LLM-Based Memory Interfaces}
\label{sec:related:llm}

The closest work applies language models to recall during or shortly after
conversation. Memoro demonstrated that short, context-aware answers delivered
through an audio wearable raise recall confidence without degrading conversation
quality, and introduced a queryless mode that infers the memory need from
ongoing context rather than waiting to be asked~\citep{zulfikar2024memoro}.
MemPal addressed object retrieval in the homes of older adults using multimodal
sensing and voice query~\citep{maniar2025mempal}, and AR Secretary Agent applied
LLM summarisation to meeting recall through augmented-reality
glasses~\citep{elhaddad2025arsecretary}. ProMemAssist advanced the timing
question that queryless assistance raises, modelling the wearer's working memory
on smart glasses in order to weigh the benefit of a proactive cue against the
cost of interrupting~\citep{pu2025promemassist}.

Two properties are shared across this group and are the ones \sysname{}
reconsiders. First, each depends on a dedicated wearable---an audio headset, a
camera unit, AR or smart glasses---which bounds deployment to users able to
obtain, wear and tolerate that device. Second, the evaluated contribution is the
interaction: what the assistant says and when. \sysname{} keeps the interaction
commitments, adopting explicit query and confidence-gated assisted recall
directly from this line, but moves the hardware requirement down to a commodity
phone or browser and adds a commitment about the \emph{content} of the answer:
that it is returned with the stored evidence it came from, and withheld when no
such evidence exists.

\subsection{Memory Architectures for LLM Agents}
\label{sec:related:agents}

A parallel literature formalises memory as a component of autonomous agents
rather than as an interface for a person. Generative Agents introduced a memory
stream retrieved by a weighted combination of recency, importance and relevance,
with periodic reflection synthesising higher-level
observations~\citep{park2023generative}; MemoryBank added a decay schedule
modelled on human forgetting together with an evolving user
portrait~\citep{zhong2024memorybank}. These architectures supply the design
vocabulary \sysname{} uses in \Cref{sec:retrieval}, and the idea that a single
similarity score is insufficient for personal memory originates here.

The limits of the current generation are equally instructive. LoCoMo evaluates
recall over very long conversations and finds that even long-context models
remain well short of human performance on questions about earlier
dialogue~\citep{maharana2024locomo}, which locates the difficulty in retrieval
rather than generation. \sysname{} differs from these systems in what memory is
\emph{for}: an agent's memory serves the agent's own behaviour and need not be
inspectable, whereas a memory assistant's store is a record of the user's life
that they must be able to read, correct and delete. That requirement, not
retrieval quality alone, is what motivates holding records, embeddings and
relations in one relational system of record rather than distributing them
across specialised stores.

\subsection{Grounding, Attribution and Abstention}
\label{sec:related:grounding}

Retrieval-augmented generation conditions a model on retrieved passages and is
the standard mechanism for tying output to evidence~\citep{lewis2020rag}. It is
not by itself sufficient. Surveys of hallucination document that models produce
fluent, well-formed statements unsupported by any source, and that retrieval
reduces without eliminating the
behaviour~\citep{huang2025hallucination}. Two responses matter here. Attribution
work such as ALCE requires the generated text to carry citations into the
retrieved evidence and evaluates whether those citations actually support the
claims~\citep{gao2023alce}. Abstention work treats declining to answer as a
first-class outcome and studies how a model can identify the boundary of what it
knows~\citep{feng2024abstain}.

\sysname{} adopts both, and the reason they become requirements rather than
refinements is specific to personal memory. In open-domain question answering a
user can, in principle, check a claim against an external source; a claim about
what was said in the user's own kitchen has no such check, and the user's own
recollection is exactly the faculty the system exists to compensate for. A
fluent invention is therefore not a degraded answer but a corrupted memory.
\sysname{} accordingly stores a provenance pointer on every record and treats
the correctness of refusals as a measured quantity alongside the correctness of
answers.

\subsection{Speech Capture and Privacy}
\label{sec:related:speech}

The front end rests on mature components: large-scale weakly supervised models
made robust transcription available across accents and noisy
conditions~\citep{radford2023whisper}, and diarization pipelines attribute
segments to speakers with usable accuracy on conversational
audio~\citep{bredin2023pyannote}. Speaker attribution is not incidental for a
memory assistant, since \emph{who} said something is often the query key.

The harder constraint is social. Studies of smart-speaker users show that
privacy concern is diffuse, poorly matched to actual system behaviour, and
managed through blunt strategies such as muting or relocating the
device~\citep{lau2018alexa}. Contextual integrity supplies the analytic frame:
what makes a disclosure appropriate is the norms of the context in which it was
made, and recent benchmarks show that language models reason poorly about those
norms even when they can recite privacy
policies~\citep{mireshghallah2024confaide}. Because a conversational memory
holds information belonging partly to people who are not its user, \sysname{}
treats visible recording state, reviewable records and reliable deletion as
design requirements rather than settings, and \Cref{sec:results} reports privacy
as the constraint participants raised first.

\subsection{Summary of Positioning}
\label{sec:related:summary}

\input{tables/tab_positioning}

\Cref{tab:positioning} summarises the comparison against the systems built for
everyday human recall. Read down the columns, every prior system in this space
requires a dedicated wearable, retrieves on a single class of signal, and is
evaluated on the interaction---what the assistant says and when---rather than on
the evidentiary status of what it says. \sysname{} occupies the remaining
position: commodity hardware, one inspectable store, retrieval that combines
four signals, and an answer that either carries its provenance or is withheld.
The agent architectures of \cref{sec:related:agents} are deliberately excluded
from the table; they supply retrieval machinery but assume a store no end user
inspects, which leaves the hardware and provenance columns inapplicable rather
than unreported.

%% file: tables/tab_positioning.tex
%% Positioning table -- non-floating, self-contained.
%% Preamble needs: tabularx, booktabs, colortbl, amssymb, caption.
\makeatletter
\@ifundefined{NC@rewrite@Y}{\newcolumntype{Y}{>{\raggedright\arraybackslash}X}}{}
\makeatother
\colorlet{rowalt}{black!4}
\colorlet{rowhead}{black!10}
\colorlet{rowours}{remoraink!9}
\newcommand{\yes}{\textcolor{remoraink}{\checkmark}}
\newcommand{\nr}{\textcolor{black!45}{--}}

\begingroup
\centering
\footnotesize
\setlength{\tabcolsep}{5pt}
\renewcommand{\arraystretch}{1.35}
\captionof{table}{Positioning of \sysname{} against memory assistants designed
  for everyday human recall. A dash (\nr) means the property is not reported by
  that work, not that the system necessarily lacks it. \emph{Provenance}: the
  answer is returned with a pointer to the stored evidence it came from.
  \emph{Abstention}: the system is specified to decline when no stored evidence
  supports the query.}
\label{tab:positioning}

\begin{tabularx}{\textwidth}{@{}>{\raggedright\arraybackslash}p{2.45cm} Y Y Y Y c c@{}}
\toprule
\rowcolor{rowhead}
\textbf{System} & \textbf{Capture} & \textbf{Hardware} &
\textbf{Memory representation} & \textbf{Retrieval} &
\textbf{Prov.} & \textbf{Abst.} \\
\midrule

Memoro\newline \citep{zulfikar2024memoro}
  & continuous audio & audio wearable & transcript $+$ embeddings
  & semantic & \nr & \nr \\

\rowcolor{rowalt}
MemPal\newline \citep{maniar2025mempal}
  & audio $+$ vision & in-home camera & object and event log
  & multimodal & \nr & \nr \\

AR Secretary\newline \citep{elhaddad2025arsecretary}
  & audio $+$ vision & AR glasses & meeting summaries
  & semantic & \nr & \nr \\

\rowcolor{rowalt}
ProMemAssist\newline \citep{pu2025promemassist}
  & multimodal sensing & smart glasses & working-memory items and episodes
  & proactive, gated by interruption cost & \nr & \nr \\

\midrule
\rowcolor{rowours}
\textbf{\sysname{}}\newline \textbf{(this work)}
  & consented speech
  & \textbf{phone or browser}
  & \textbf{records, embeddings and relations in one store}
  & \textbf{lexical, semantic, relational, recency}
  & \yes & \yes \\
\bottomrule
\end{tabularx}
\par\endgroup
\vspace{0.9\baselineskip}

%% file: sections/03_design_goals.tex
\section{Design Goals and Methodology}
\label{sec:goals}

The design of \sysname{} is governed by five goals. Each is derived either from
the requirements survey reported in \cref{sec:evalmethod} or from the three
constraints established in \cref{sec:introduction}, and each is stated so that a
violation is observable rather than a matter of judgement. We give the goal, the
reason it binds for this population, and the condition under which it would be
considered unmet.

\subsection{Design Goals}
\label{sec:goals:list}

\begin{description}[leftmargin=0pt,style=unboxed,itemsep=0.5em,topsep=0.4em]

  \item[\textbf{G1 --- Minimal input.}] A short free-form question, spoken or
    typed, must be sufficient to retrieve a memory. The user should not have to
    learn a command syntax, fill fields, or phrase the query in the words the
    system happens to have stored. This follows directly from the argument in
    \cref{sec:introduction}: any input cost is levied on attention that the
    surrounding activity has already claimed. \emph{Unmet if} successful recall
    depends on the user guessing the stored wording, or if queries routinely
    require reformulation.

  \item[\textbf{G2 --- Minimal output.}] An answer must be short enough to be
    consumed while the user is doing something else---the name, the phrase, the
    date---rather than a paragraph containing it. Cognitive load theory gives
    the reason~\citep{sweller2024cognitive}: a long answer delivered during an
    ongoing task substitutes one load for another and can leave the user no
    better off than the lapse did. \emph{Unmet if} answers must be re-read or
    re-heard to extract the operative detail.

  \item[\textbf{G3 --- Grounded recall.}] An answer must be produced from
    retrieved memory records and returned with a pointer to the record that
    supports it; where no record supports the query, the system must say so
    rather than produce a plausible sentence. Attribution and abstention are
    treated as a single commitment because either alone is
    insufficient~\citep{gao2023alce,feng2024abstain}. \emph{Unmet if} any answer
    is returned that no stored record entails.

  \item[\textbf{G4 --- User control and privacy.}] Recording state must be
    visible whenever capture is active; stored records must be reviewable and
    editable; and deletion must remove the record together with its embeddings,
    tags, relationships and answer citations. This goal is set higher than
    convention because the store holds speech belonging partly to people who are
    not its user, and because smart-speaker research shows that users manage
    unclear system behaviour with blunt instruments---muting, unplugging,
    relocating the device---rather than with the controls
    provided~\citep{lau2018alexa}. \emph{Unmet if} deletion is cosmetic, that
    is, if any derived artefact of a deleted record remains retrievable.

  \item[\textbf{G5 --- Practical deployment.}] The system must run on hardware
    the intended users already own. Web and mobile clients are the primary
    interface; wearable output is supported but never required. \emph{Unmet if}
    any core capability is reachable only through a dedicated device.

\end{description}

\subsection{Traceability}
\label{sec:goals:traceability}

\input{tables/tab_traceability}

The goals are not independent. G1 and G2 jointly define the interaction budget,
and both are spent by the same resource; G3 and G4 both concern what the user is
entitled to know about the system's output and its store. G5 constrains the
others by removing the option of solving an interaction problem with dedicated
hardware---the solution most of the systems in \cref{tab:positioning} adopt.

\subsection{Iterative Development Method}
\label{sec:goals:method}

\sysname{} was built in vertical slices, each ending in a capability usable
end to end rather than in a component awaiting integration: session capture,
then transcription, then memory extraction, then embedding and storage, then
retrieval, then grounded answer generation, then assisted recall. Building in
this order meant that the interaction goals G1 and G2 could be exercised against
a working path from the third slice onwards, rather than being deferred until
the retrieval stack was complete.

The most consequential change during development was architectural, and it
followed from G4 rather than from any performance consideration. Early versions
held relational records in PostgreSQL, vectors in a standalone vector store, and
relationships in a dedicated graph database. That arrangement was workable while
memories were only ever appended, and became unmanageable once editing and
deletion were implemented: a single deletion had to succeed across three systems
with no shared transaction, and any partial failure left a deleted memory
retrievable through one of the remaining indexes. Since G4 makes cosmetic
deletion a failure condition rather than a defect to be tolerated, the stores
were consolidated into one PostgreSQL instance using \texttt{pgvector} for
similarity, full-text search for lexical matching, relationship tables for
bounded graph traversal, and PostGIS where location is relevant. Consolidation
was adopted for delete and revision consistency; the ability to combine all four
retrieval signals within a single query plan, described in
\cref{sec:retrieval}, was a consequence rather than the motivation.

%% file: tables/tab_traceability.tex
%% Goal -> mechanism -> measure traceability. Non-floating, self-contained.
\makeatletter
\@ifundefined{NC@rewrite@Y}{\newcolumntype{Y}{>{\raggedright\arraybackslash}X}}{}
\makeatother
\begingroup
\colorlet{trowalt}{black!4}
\colorlet{trowhead}{black!10}
\centering
\footnotesize
\setlength{\tabcolsep}{5pt}
\renewcommand{\arraystretch}{1.3}
\captionof{table}{Each design goal, the mechanism that implements it, and the
  measure by which its violation would be observed. \Cref{sec:evalmethod}
  defines the measures; \cref{sec:results} reports against them.}
\label{tab:traceability}

\begin{tabularx}{\textwidth}{@{}>{\raggedright\arraybackslash}p{2.9cm} Y Y@{}}
\toprule
\rowcolor{trowhead}
\textbf{Goal} & \textbf{Mechanism} & \textbf{Observable measure} \\
\midrule
\textbf{G1} Minimal input
  & short free-form voice or text query; no command syntax or form fields
  & interaction cost: query length and reformulations per successful recall \\
\rowcolor{trowalt}
\textbf{G2} Minimal output
  & conciseness pass constraining the answer to the span that satisfies the need
  & response length, and perceived load in the post-task questionnaire \\
\textbf{G3} Grounded recall
  & generation restricted to retrieved records; explicit refusal when none supports the query
  & answer faithfulness, and correctness of refusals \\
\rowcolor{trowalt}
\textbf{G4} User control
  & visible recording state; records reviewable and editable; deletion cascading to embeddings, tags, relations and citations
  & privacy comfort, and whether deleted content remains retrievable \\
\textbf{G5} Practical deployment
  & web and mobile clients on commodity devices; wearable output optional
  & whether sessions complete on participants' own hardware \\
\bottomrule
\end{tabularx}
\par\endgroup
\vspace{0.9\baselineskip}

%% file: sections/04_system_design.tex
\section{System Design}
\label{sec:design}

\begin{figure}[htbp]
  \centering
  \includegraphics[width=\textwidth]{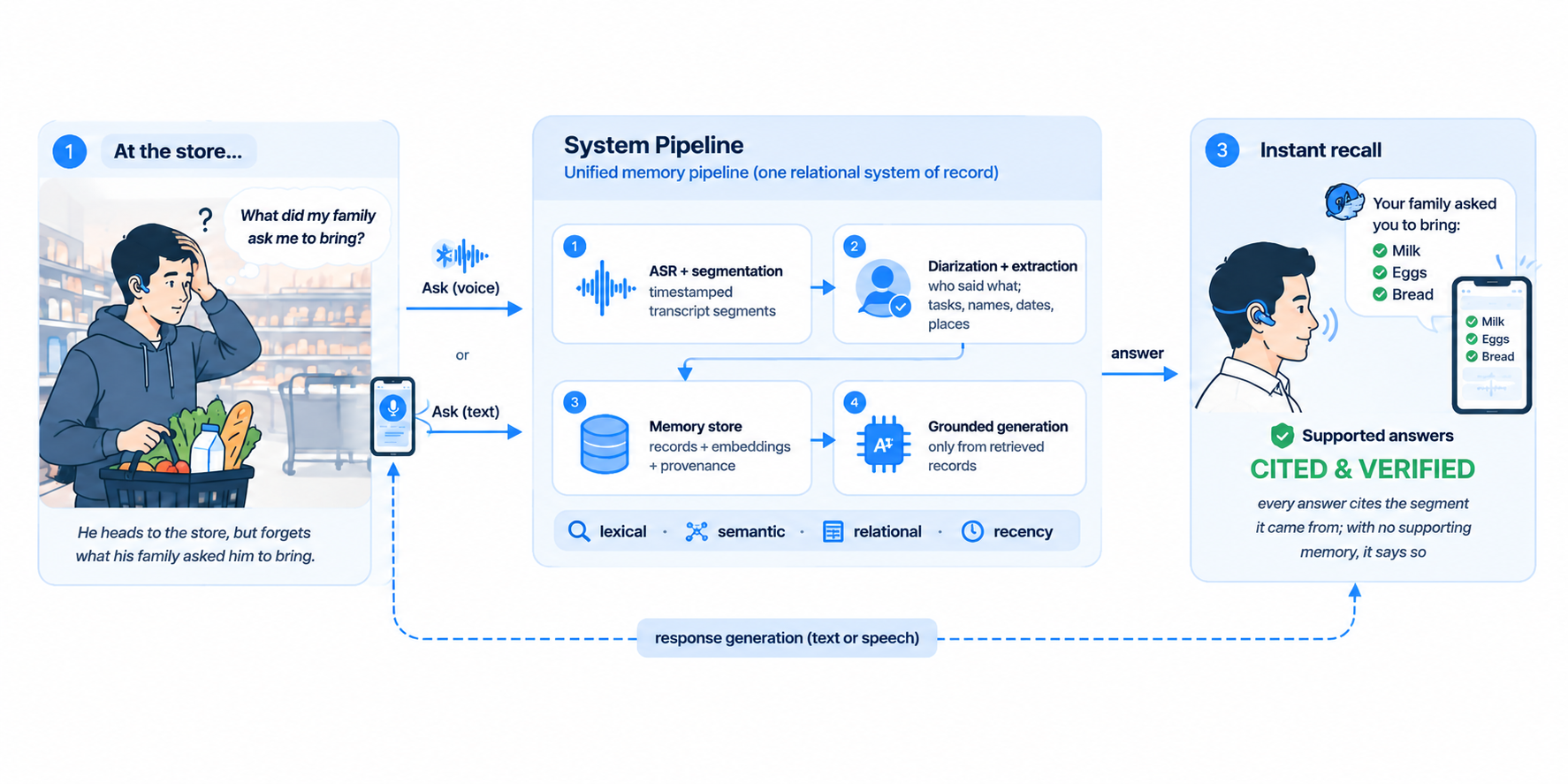}
  \caption{\sysname{} system overview. A recall need arises away from the moment
    of capture and is expressed as a short voice or text query. The unified
    memory pipeline transcribes and segments captured speech, attributes
    segments to speakers and extracts memory candidates, stores each as a record
    carrying provenance, and answers from retrieved records only. Retrieval
    combines four signals over a single relational system of record, and the
    answer is returned with the segment it came from.}
  \label{fig:overview}
\end{figure}

\sysname{} is organised as two pipelines over one store. A \emph{logging}
pipeline converts captured speech into structured memory records; a
\emph{recall} pipeline answers a query from those records. The two are
deliberately asymmetric. Logging is asynchronous and may take seconds per
minute of audio, because nothing waits on it; recall is synchronous and
latency-bound, because a user is standing in a shop aisle. Placing transcription,
diarization, extraction and embedding entirely in the logging path is what makes
the recall path short enough to be worth using, and is the reason the store
holds derived artefacts rather than computing them on demand.
\Cref{fig:overview} shows both paths.

\subsection{Memory Encoder}
\label{sec:design:encoder}

The encoder turns raw audio into candidate memories. Audio is captured through
the web or mobile client, uploaded to object storage, and enqueued for
background processing; the client's only synchronous responsibility is to make
the recording state visible and to bind the capture to a session, as G4
requires.

Processing proceeds in three stages. \textbf{Transcription} produces text with
word-level timings. \textbf{Diarization} attributes stretches of that text to
distinct speakers, which matters more for a memory assistant than for
transcription generally: \emph{who} said something is frequently the query key,
and an instruction from a clinician and a remark from a passer-by have different
standing even when their wording is identical. \textbf{Segmentation} then
divides the transcript at speaker and topic boundaries into units small enough
to be retrieved independently but large enough to remain interpretable in
isolation.

A segment becomes a \emph{memory candidate} when it carries information that may
be needed after the conversation ends. In practice this means it contains at
least one of: a commitment or task; an appointment, date or deadline; an
instruction with a condition attached; a named person, place or object; the
stated location of an object; or a decision. Segments that carry only
conversational scaffolding---greetings, acknowledgements, repairs---are
discarded rather than stored, since retaining them dilutes retrieval without
adding recall value. Candidate identification and field extraction are performed
by a language model against a fixed schema, and the extracted summary is stored
alongside, never in place of, the source text.

\subsection{Memory Record Schema}
\label{sec:design:record}

Each memory candidate is stored as a record
\begin{equation}
  \mem_i = \{\, t_i,\; s_i,\; x_i,\; a_i,\; \emb{i},\; z_i,\; c_i \,\},
  \label{eq:record}
\end{equation}
where $t_i$ is the timestamp of the underlying utterance, $s_i$ the speaker
label, $x_i$ the verbatim source segment, $a_i$ the extracted summary,
$\emb{i}$ the embedding of $a_i$, $z_i$ the extracted entities, tags and task
fields, and $c_i$ a citation pointer to the transcript segment from which the
record was derived.

The fields divide by the role they play. $x_i$ and $a_i$ are two views of the
same content and both are kept: $a_i$ is what the retriever embeds and what the
generator reads, while $x_i$ is what the user is shown when they ask what a
memory is based on. Storing only the summary would make the system unable to
answer that question, and a memory assistant that cannot show its working is in
the position the paraphrase problem describes---the user must take the summary
on trust. $\emb{i}$ and $z_i$ feed the semantic and relational retrieval signals
respectively. $t_i$ supports recency ranking and the many queries that are
temporal rather than topical (``what did she say yesterday'').

The field that carries the paper's central commitment is $c_i$. Because every
record retains a pointer back into the session that produced it, an answer can
be returned with the conversation and the transcript it came from, and the user can verify the system against a source rather than
against their own recollection---the faculty the system exists to compensate
for. $c_i$ is also what makes abstention well defined: if retrieval returns no
record, there is no $c_i$ to cite, and the absence of a citation is the
condition under which the generator must decline rather than a heuristic applied
afterwards.

\subsection{Structured Memory Store}
\label{sec:design:store}

The store holds sessions, captures, transcript segments, memory records, record
revisions, entities, relationships between records and entities, tags, answers
and answer citations. All of it resides in a single PostgreSQL instance, with
\texttt{pgvector} for similarity, full-text search for lexical matching,
relationship tables for bounded traversal, and PostGIS where a memory is tied to
a place. \Cref{sec:goals:method} gave the reason this is one store rather than
three: deletion and revision must be consistent, and a deletion spanning
separate systems has no shared transaction to make it so.

Two consequences follow for the design. First, deletion is a cascade rather than
a flag. Removing a record removes its embedding, its tags, its relationship
edges, and any answer citations that reference it, in one transaction; a
deleted memory therefore cannot resurface through a stale index, which is the
failure mode G4 rules out. Second, revisions are tracked rather than overwritten,
so that a user who corrects a misheard name changes what the system will answer
without destroying the evidence of what was originally captured. A memory
assistant is a record of the user's life, and silently rewriting such a record
is a different and worse failure than getting a transcription wrong in the first
place.

The store is also what makes the retrieval formulation of \cref{sec:retrieval}
expressible. Because metadata filtering, vector distance, relationship joins and
recency all live in one system, a single query plan can apply them together
rather than reconciling ranked lists returned by separate services---an
arrangement that, as \cref{tab:positioning} notes, none of the comparable
systems reports.

%% file: sections/05_retrieval_generation.tex
\section{Hybrid Retrieval and Grounded Generation}
\label{sec:retrieval}

This section specifies how a query becomes an answer. It is where the two
commitments made in \cref{sec:goals}---grounded recall (G3) and an interaction
budget (G1, G2)---are discharged, and where \sysname{} differs most sharply
from a general assistant pointed at a transcript archive.

\subsection{Why a Single Retrieval Signal Is Insufficient}
\label{sec:retrieval:why}

Personal memory queries are not well served by any one retrieval method, and the
reason is that they do not form a single class. \Cref{tab:queryclasses} sets out
the classes observed during the pilot together with the signal that resolves
each and the signal that fails when used alone.

\input{tables/tab_queryclasses}

Two observations follow. First, the failures are not marginal: a semantic
retriever confronted with ``what was the tablet called'' has almost nothing to
work with, because the informative token is a proper noun that embeddings place
poorly, while a lexical retriever confronted with a paraphrase has no overlap at
all. Each method fails completely on the class the other handles. Second, two of
the six classes carry almost no topical content. ``What did she say yesterday''
is answered by a speaker and a time, not by a topic, and a system that ranks
only on content similarity has no way to serve it. Hybrid retrieval is therefore
not an incremental gain over the best single signal; it is what makes several
common query classes answerable at all.

The final row of \cref{tab:queryclasses} is of a different kind. When the user
asks about something never captured, every signal will still return its
best-ranked candidates, because ranking functions are not designed to return
nothing. Handling that row is the job of the grounding gate in
\cref{sec:retrieval:generation}, not of the retriever.

\subsection{Retrieval Signals and Their Fusion}
\label{sec:retrieval:signals}

\sysname{} computes several ranked lists over the memory store and fuses them
into one. Lexical relevance $\lex$ comes from full-text search over the summary
$a_i$ and the source segment $x_i$, and resolves exact identifiers, proper nouns
and numbers. Semantic relevance $\vecsim$ is cosine similarity between the query
embedding and the record embedding,
\begin{equation}
  \vecsim(\mem_i,\query) = \frac{\emb{i} \cdot \emb{\query}}
    {\lVert \emb{i} \rVert \, \lVert \emb{\query} \rVert},
  \label{eq:cosine}
\end{equation}
and absorbs paraphrase. Relational relevance $\ctx$ traverses the entity and
relationship tables from entities named in the query to the records mentioning
them, making person- and object-anchored queries tractable. Proximity $\prox$
contributes recency and, where a memory is tied to a place, spatial nearness
through PostGIS.

These signals are not commensurable: a BM25-style lexical score, a cosine
similarity in $[-1,1]$ and a graph-traversal score have different ranges and
different distributions, so summing them requires a normalisation that is itself
a tuned object. \sysname{} therefore fuses by \emph{rank} rather than by score,
using reciprocal rank fusion~\citep{cormack2009rrf}. Writing $r_s(\mem_i)$ for
the rank of record $\mem_i$ in the list produced by signal $s$, and $S$ for the
set of signals returning that record,
\begin{equation}
  \score(\mem_i, \query) \;=\;
    \sum_{s \in S} \frac{1}{\kappa + r_s(\mem_i)},
  \qquad \kappa = 60 ,
  \label{eq:hybrid}
\end{equation}
and the top $k$ records by $\score$ are passed forward. Rank fusion has two
properties that matter here. It requires no per-signal weights, so there are no
parameters to overfit to a small corpus; and because the contribution of a
signal decays with rank rather than with score, a record ranked highly by one
signal survives fusion even when the other signals do not retrieve it at all,
which is exactly the behaviour the single-signal failures of
\cref{tab:queryclasses} demand. The constant $\kappa$ damps the influence of
top-ranked items from any one list; we use the conventional value of 60 and did
not tune it.

Because all signals are computed within one PostgreSQL instance
(\cref{sec:design:store}), \cref{eq:hybrid} is evaluated over one set of rows
rather than by reconciling results returned by separate services---which is what
makes rank fusion straightforward to apply, since every list is produced by the
same engine over the same identifiers.

\subsection{Query Mode and Assisted Recall}
\label{sec:retrieval:modes}

Retrieval is reached by two routes that differ in who initiates and, critically,
in what a mistake costs.

In \textbf{Query Mode} the user asks a short question. A false positive here is
cheap: the user sees an answer that does not help, discards it, and asks again.
The design objective is coverage, and the threshold for producing an answer is
the grounding gate alone.

In \textbf{Assisted Recall} the system infers from recent context that a memory
may be wanted and surfaces a cue unasked. A false positive is now expensive,
because it interrupts a conversation that was proceeding without difficulty, and
repeated interruption is the failure that makes proactive assistants get
switched off. \sysname{} therefore requires three conditions to hold before an
unsolicited cue is surfaced, and stays silent if any fails:

\begin{enumerate}[leftmargin=1.6em,itemsep=0.25em,topsep=0.3em]
  \item \textbf{A need is inferable.} The context must yield a well-formed
    retrieval query. Silence, filler and low-information stretches yield none,
    and are discarded before retrieval is attempted.
  \item \textbf{Retrieval returns usable evidence.} The inferred query must
    retrieve records at all; an empty result suppresses the cue. Because the
    cost of a false positive is higher here than in Query Mode, this gate is
    intended to be stricter than the one applied to explicit queries, and
    \cref{sec:limitations} notes that quantifying that asymmetry requires the
    retrieval measurements this paper does not yet report.
  \item \textbf{The cue is not redundant or sensitive.} A cue repeating what was
    just said is suppressed, as is one that would speak aloud a memory
    attributable to a third party. The second condition exists because an
    unsolicited answer may be heard by people who are not the user, which is the
    contextual-integrity problem of
    \cref{sec:related:speech}~\citep{mireshghallah2024confaide}.
\end{enumerate}

Silence is thus the default rather than the fallback: assisted recall speaks
only when all three gates pass, and the system is designed to under-trigger.

\subsection{Grounded Generation and Abstention}
\label{sec:retrieval:generation}

Only the top-ranked records reach the generator, and the generator is
constrained to answer from them. The contract has three clauses. \emph{Answer
only from the retrieved records}: no background knowledge, no inference beyond
what the records state. \emph{Return the citation}: the answer carries the $c_i$
of the records used, so the user can see the conversation and moment it came
from. \emph{Otherwise decline}: if retrieval returns nothing above threshold, or
if the generated answer is not entailed by the retrieved records, the system
reports that the memory is not available.

Abstention is anchored to the retrieved evidence rather than to the model's
confidence in its own knowledge: the generator is given the retrieved records
and instructed to answer from them or report their insufficiency, so the
judgement it makes is about a set of documents in front of it rather than about
the boundary of what it knows. This is a weaker guarantee than a retrieval-side
score threshold, which would make abstention independent of the generator
altogether; \cref{sec:limitations} identifies that threshold as the main
outstanding strengthening of the contract. The prompts implementing the contract are given verbatim in
\cref{app:prompts}, and the full procedure is \cref{alg:recall}.

\input{algorithms/alg_logging_recall}

A concise-rewrite pass follows generation, reducing the answer to the span that
satisfies the need. This is G2, and it is placed after generation rather than
folded into it so that conciseness cannot come at the cost of grounding: the
rewrite operates on text already constrained to the evidence, and cannot
introduce content the records do not support.

Two properties of this arrangement are worth stating because they are what
\cref{sec:evalmethod} sets out to measure. Abstention is a first-class outcome
with its own correctness, not a degraded answer: a refusal when the memory
exists and a fabrication when it does not are distinct failures with distinct
costs, and both must be counted. And every non-abstaining answer is auditable,
because the citation identifies the evidence a reader can check it against.

%% file: tables/tab_queryclasses.tex
%% Query classes and the signal each one defeats. Non-floating, self-contained.
\makeatletter
\@ifundefined{NC@rewrite@Y}{\newcolumntype{Y}{>{\raggedright\arraybackslash}X}}{}
\makeatother
\begingroup
\colorlet{qrowalt}{black!4}
\colorlet{qrowhead}{black!10}
\centering
\footnotesize
\setlength{\tabcolsep}{5pt}
\renewcommand{\arraystretch}{1.3}
\captionof{table}{Query classes observed in the pilot, the signal that resolves
  each, and the signal that fails when used alone. The final row is not a
  retrieval case: no signal should resolve it, and the required behaviour is
  abstention.}
\label{tab:queryclasses}

\begin{tabularx}{\textwidth}{@{}>{\raggedright\arraybackslash}p{2.3cm} Y >{\raggedright\arraybackslash}p{2.1cm} Y@{}}
\toprule
\rowcolor{qrowhead}
\textbf{Query class} & \textbf{Example} & \textbf{Resolved by} &
\textbf{Fails alone} \\
\midrule
Exact identifier
  & ``What was the tablet called?''
  & lexical
  & semantic: rare proper nouns sit far from their neighbours in embedding space \\
\rowcolor{qrowalt}
Paraphrase
  & user asks about ``tablets''; the record says ``medicine''
  & semantic
  & lexical: no term overlap \\
Person-anchored
  & ``What task did Maaz assign me?''
  & relational
  & lexical and semantic: the topic is underspecified, the speaker is not \\
\rowcolor{qrowalt}
Temporal
  & ``What did she say yesterday?''
  & recency
  & lexical and semantic: the query carries almost no topical content \\
Place-anchored
  & ``Where did I keep my documents?''
  & relational and location
  & semantic: many records mention documents \\
\rowcolor{qrowalt}
\textit{Unsupported}
  & ``What did we decide about the lease?'' when no such conversation was captured
  & \textit{none}
  & \textit{all}: any non-empty result here is a false positive \\
\bottomrule
\end{tabularx}
\par\endgroup
\vspace{0.9\baselineskip}

%% file: algorithms/alg_logging_recall.tex
\begin{algorithm}[htbp]
\caption{\sysname{} recall: hybrid retrieval with grounded generation and
  abstention}
\label{alg:recall}
\begin{algorithmic}[1]
\Require query $\query$ (explicit) or recent context $\ctx$ (assisted);
         memory store $\memset$; threshold $\tau$; cut-off $k$
\Ensure  a cited answer $Y$, an explicit refusal, or silence
\Statex
\Statex \textbf{\textit{// 1. Obtain a query}}
\If{the user asks explicitly}
  \State $\query \gets \Call{NormaliseOrTranscribe}{\query}$
\Else \Comment{assisted recall}
  \State $\query \gets \Call{InferMemoryNeed}{\ctx}$
  \If{$\query = \varnothing$} \State \Return silence \EndIf
\EndIf
\Statex
\Statex \textbf{\textit{// 2. Score every candidate on four signals}}
\ForAll{$\mem_i \in \Call{CandidateSet}{\query}$}
  \State $\score(\mem_i,\query) \gets
          \alpha\lex + \beta\vecsim + \gamma\ctx' + \delta\prox$
\EndFor
\State $\topk \gets \Call{TopK}{\score, k}$
\State $\topk \gets \Call{Deduplicate}{\topk}$
  \Comment{collapse revisions of the same record}
\Statex
\Statex \textbf{\textit{// 3. Grounding gate --- evidence decides, not the model}}
\If{$\topk = \varnothing$ \textbf{ or } $\max \score < \tau$}
  \State \Return ``that memory is not available'' \Comment{abstain}
\EndIf
\Statex
\Statex \textbf{\textit{// 4. Generate, constrained to the retrieved evidence}}
\State $Y \gets \Call{GenerateFrom}{\query, \topk}$
\State $Y \gets \Call{Shorten}{Y}$ \Comment{G2: sized for speech}
\If{$Y$ is not entailed by $\topk$}
  \State \Return ``that memory is not available''
\EndIf
\State attach citations $\{c_i : \mem_i \in \topk \text{ used by } Y\}$
\Statex
\Statex \textbf{\textit{// 5. Privacy gate applies to unsolicited cues only}}
\If{assisted \textbf{ and } \textbf{not} $\Call{PrivacyGate}{\query, \topk}$}
  \State \Return silence
\EndIf
\State \Return $Y$ with citations
\end{algorithmic}
\end{algorithm}

%% file: sections/06_implementation.tex
\section{Implementation}
\label{sec:implementation}

\sysname{} is implemented as a web and mobile client over a background worker
and a single relational system of record. This section reports what was built,
how partial failures are handled, and why each major component was chosen or
discarded. Exact model identifiers, embedding dimensionality, retrieval
parameters and service versions are collected in \cref{app:repro}; the
interface is shown in \cref{app:interface}.

\subsection{Architecture}
\label{sec:impl:stack}

\input{tables/tab_components}

The division of work follows the asymmetry described in \cref{sec:design}. The
clients do nothing expensive: they bind a capture to a session, make the
recording state visible, and upload. Everything costly---transcription,
diarization, extraction, embedding---runs in the worker, off the path the user
waits on. The API layer is thin by design, because the operations that matter
for correctness are database transactions rather than application logic, and
placing them in the database is what makes the deletion guarantee of G4
enforceable rather than aspirational.

\subsection{Processing States and Failure Handling}
\label{sec:impl:states}

A capture moves through an explicit state machine: \textsf{created},
\textsf{uploaded}, \textsf{queued}, \textsf{processing}, \textsf{processed}, or
\textsf{failed}. The states are recorded rather than inferred, for a reason
specific to this application. Each stage of the encoder calls an external
service, and external calls fail: a transcription request times out, a rate
limit is hit, an extraction returns malformed output. A pipeline without
explicit states has two bad options on retry---skip the capture, losing memories
the user believes were recorded, or reprocess it, producing duplicate records
that then compete with each other in retrieval. Explicit states make retries
idempotent: a capture is resumed from its last completed stage, and a record is
written once.

Failure is therefore visible rather than silent. A capture stuck in
\textsf{failed} is surfaced in the client, because the alternative---a user who
believes a conversation was captured when it was not---is the worst failure mode
available to a memory assistant. It converts a system that sometimes cannot
answer into a system the user cannot calibrate against, and calibration is what
makes an assistive tool usable at all.

Deletion is implemented as a cascade within one transaction, as
\cref{sec:design:store} requires: removing a record removes its embedding, tags,
relationship edges and answer citations together, so no derived artefact can
outlive the record it was derived from. Edits create a revision rather than
overwriting, and retrieval deduplicates revisions of the same record so that a
corrected memory does not compete with its own earlier version.

\subsection{Model and Tool Selection}
\label{sec:impl:selection}

\input{tables/tab_model_selection}

Three decisions in \cref{tab:selection} are worth drawing out. Transcription was
moved to a hosted service despite a working local pipeline, because every
downstream stage consumes the transcript and an error there propagates into the
stored record, the embedding and ultimately the answer; accuracy was judged the
binding constraint and provider independence the thing to trade. Generation uses
a compact model rather than a large-context one, which is a consequence of the
grounding contract: only the top-ranked records are passed, so the context is
small by construction and long-context capacity buys nothing. And the two
tinted rows record the consolidation described in \cref{sec:goals:method}---a
standalone vector database and a graph database were both removed, not because
they retrieved poorly, but because neither could participate in the transaction
that a deletion has to be.

\subsection{Interface}
\label{sec:impl:ui}

The client surfaces are shown in \cref{app:interface}. Two of them carry design
commitments rather than convenience. The capture view displays recording state
continuously while audio is being taken, which is the visible half of G4 and the
signal a bystander can act on. The answer view displays the citation alongside
the answer, so that the provenance stored in $c_i$ is reachable in one
interaction rather than buried in a detail screen; an answer whose evidence is
three taps away is, in practice, an answer without evidence.

%% file: tables/tab_components.tex
%% Component table. Non-floating, self-contained.
\makeatletter
\@ifundefined{NC@rewrite@Y}{\newcolumntype{Y}{>{\raggedright\arraybackslash}X}}{}
\makeatother
\begingroup
\colorlet{crowalt}{black!4}
\colorlet{crowhead}{black!10}
\centering
\footnotesize
\setlength{\tabcolsep}{5pt}
\renewcommand{\arraystretch}{1.28}
\captionof{table}{\sysname{} components. Version identifiers and service
  endpoints are listed in \cref{app:repro} rather than cited here, to keep the
  reference list to scholarly sources.}
\label{tab:components}

\begin{tabularx}{\textwidth}{@{}>{\raggedright\arraybackslash}p{2.6cm} >{\raggedright\arraybackslash}p{3.1cm} Y@{}}
\toprule
\rowcolor{crowhead}
\textbf{Layer} & \textbf{Technology} & \textbf{Responsibility} \\
\midrule
Web client & Next.js
  & capture with visible recording state, memory timeline, query entry, answer and citation display \\
\rowcolor{crowalt}
Mobile client & Expo / React Native
  & capture and recall away from a desk, which is where most recall needs arise \\
API & Bun $+$ Hono
  & session lifecycle, query endpoints, access control \\
\rowcolor{crowalt}
Worker & Bun background jobs
  & transcription, diarization, extraction, embedding; retries and state transitions \\
Authentication & Supabase Auth
  & single-owner access; every record is scoped to one account \\
\rowcolor{crowalt}
System of record & Supabase PostgreSQL
  & sessions, segments, records, revisions, entities, relationships, answers, citations \\
Semantic index & \texttt{pgvector}
  & embedding similarity, $\vecsim$ in \cref{eq:hybrid} \\
\rowcolor{crowalt}
Lexical index & PostgreSQL full-text search
  & term and phrase matching, $\lex$ in \cref{eq:hybrid} \\
Spatial index & PostGIS
  & proximity for place-anchored recall, part of $\prox$ \\
\rowcolor{crowalt}
AI services & speech and language model APIs
  & transcription, extraction, grounded generation, optional speech output \\
\bottomrule
\end{tabularx}
\par\endgroup
\vspace{0.9\baselineskip}

%% file: tables/tab_model_selection.tex
%% Model/tool selection rationale. Non-floating, self-contained.
\makeatletter
\@ifundefined{NC@rewrite@Y}{\newcolumntype{Y}{>{\raggedright\arraybackslash}X}}{}
\makeatother
\begingroup
\colorlet{mrowalt}{black!4}
\colorlet{mrowhead}{black!10}
\colorlet{mrowdrop}{remoraink!8}
\centering
\footnotesize
\setlength{\tabcolsep}{5pt}
\renewcommand{\arraystretch}{1.28}
\captionof{table}{Options evaluated during development and the reason each was
  kept or dropped. The two tinted rows were removed for the consistency reason
  given in \cref{sec:goals:method}, not for retrieval quality.}
\label{tab:selection}

\begin{tabularx}{\textwidth}{@{}>{\raggedright\arraybackslash}p{2.4cm} >{\raggedright\arraybackslash}p{2.7cm} Y >{\raggedright\arraybackslash}p{1.6cm}@{}}
\toprule
\rowcolor{mrowhead}
\textbf{Role} & \textbf{Option} & \textbf{Reason for the decision} &
\textbf{Outcome} \\
\midrule
Transcription & Faster-Whisper (local)
  & runs offline on modest hardware; accuracy degraded on noisy and
    code-switched speech, which is common in the deployment setting
  & tested \\
\rowcolor{mrowalt}
Transcription $+$ diarization & WhisperX / pyannote pipeline
  & good alignment and speaker attribution; heavier to deploy and operate
  & tested \\
Transcription & hosted speech API
  & recall quality depends directly on transcript quality, so accuracy was
    weighted above independence from a provider
  & selected \\
\rowcolor{mrowalt}
Generation & compact instruction-following model
  & only the retrieved records are passed, so long-context capacity is not the
    binding constraint; cost and latency are
  & selected \\
Embedding & hosted text embedding model
  & semantic matching carries the paraphrase class in \cref{tab:queryclasses}
  & selected \\
\rowcolor{mrowdrop}
Vector store & standalone vector database
  & separate lifecycle from the relational rows; deletions could not be made
    atomic with them
  & \textbf{dropped} \\
\rowcolor{mrowdrop}
Relationship store & dedicated graph database
  & a third system in the delete path, for traversals shallow enough to express
    as SQL joins
  & \textbf{dropped} \\
Everything above & one PostgreSQL instance with extensions
  & one transaction boundary; all four retrieval signals in one query plan
  & selected \\
\rowcolor{mrowalt}
Speech output & text-to-speech, optional earpiece
  & hands-free recall where it helps, never required (G5); audible answers are
    also a privacy exposure in public
  & optional \\
\bottomrule
\end{tabularx}
\par\endgroup
\vspace{0.9\baselineskip}

%% file: sections/07_evaluation_method.tex
\section{Evaluation Methodology}
\label{sec:evalmethod}

\sysname{} is evaluated along three lines, each answering a different question.
A requirements survey establishes whether the problem is real and what
constrains adoption. A component evaluation on public speech corpora
establishes how the capture stage behaves under the acoustic conditions the
system will meet. A purpose-built recall corpus specifies how retrieval,
grounding and abstention are to be measured. The first two are reported in
\cref{sec:results}; the third is released with the paper and its measurement
protocol is given here.

\subsection{Study 1: Requirements Survey}
\label{sec:evalmethod:survey}

Fifty participants across a range of age groups completed a ten-item
questionnaire on how often conversational detail is forgotten, what it costs
them, how they currently compensate, and what they would accept from a
memory-assisting system. The instrument is reproduced in \cref{app:survey}. The
survey was designed to validate the problem and to elicit adoption constraints;
it establishes nothing about system performance, and we draw no performance
claim from it.

\subsection{Study 2: Pilot Feasibility Test}
\label{sec:evalmethod:pilot}

Ten participants with self-reported memory difficulty used \sysname{} in
controlled, semi-realistic scenarios: a short conversation was captured, and the
participant later asked for a detail from it. The procedure had five stages
(introduction and consent, demonstration, scenario capture, recall interaction,
post-task questionnaire) and the scenarios covered conversation, task,
appointment, name and object-location recall, together with one query about
information that had never been captured. The questionnaire is in
\cref{app:questionnaire} and the scenarios in \cref{app:scenarios}. With ten
participants the study supports description, not inference, and we report it
descriptively throughout.

\subsection{Transcription Evaluation}
\label{sec:evalmethod:asr}

The capture stage was evaluated on public speech corpora carrying published
reference transcripts, so that the measurement is reproducible and comparable
rather than specific to audio we recorded ourselves. We used LibriSpeech
\textsf{test-clean} and \textsf{test-other}, and the MUCS 2021 Hindi--English
code-switched test set, with noise drawn from MUSAN.

Two augmentations were applied deterministically to \textsf{test-clean}:
additive noise scaled so that the full-clip RMS of the noise meets the speech
RMS at 20, 10 and 5\,dB SNR, and a synthetic reverberation with an
exponentially decaying impulse response (RT60 $=0.35$\,s, wet mix 0.6). No audio
was recorded for this study and none was synthesised, so no text-to-speech
output enters any reported figure. Word error rate is computed by word-level
Levenshtein alignment after a standard normaliser (lowercase, punctuation
stripped, numerals expanded), reporting substitutions, deletions and insertions
separately.

\subsection{Recall Corpus and Measurement Protocol}
\label{sec:evalmethod:corpus}

Measuring retrieval, grounding and abstention requires knowing what was said and
what should be recoverable from it. We therefore constructed a corpus of 25
conversations---clinic visits, family conversations, team meetings, lectures,
shop interactions and phone calls---in clean, noisy (overlapping speech, false
starts, self-corrections) and code-switched (Roman-script Urdu mixed with
English, as spoken in Karachi) conditions. Each conversation is annotated with
its segments, the memories that should be extracted from it, and a set of
queries, each carrying the identifier of the segment that supports its answer.

\input{tables/tab_corpus}

\Cref{tab:corpus} gives the composition. Two properties of its construction
matter. First, 28\% of queries are \emph{unsupported}: they ask about
information never present in that conversation. Without them, abstention cannot
be measured at all, since a corpus of answerable questions can only reward a
system for answering. Second, the corpus was subjected to an automated
difficulty audit before use, verifying that paraphrase queries share little
vocabulary with their supporting segment, that no unsupported query names an
entity occurring anywhere in its conversation, that exact-identifier queries
turn on a token appearing in at most one segment, and that temporal queries are
not resolvable by content similarity alone. \Cref{sec:results:corpus} reports
what the audit revealed.

\subsection{Measures}
\label{sec:evalmethod:measures}

Against the goals of \cref{tab:traceability}, the protocol defines the
following.

\textbf{Transcription quality} is word error rate,
\begin{equation}
  \mathrm{WER} = \frac{S + D + I}{N},
  \label{eq:wer}
\end{equation}
with $S$ substitutions, $D$ deletions, $I$ insertions and $N$ reference words.

\textbf{Retrieval quality} is measured per query class as Recall@$k$ and
Precision@$k$,
\begin{equation}
  \mathrm{Precision}@k =
    \frac{\lvert \text{relevant records in } \topk \rvert}{k},
  \label{eq:patk}
\end{equation}
together with mean reciprocal rank, where a record counts as relevant when its
provenance identifies a supporting segment of the query. Retrieval is evaluated
in three configurations---combined, lexical only and semantic only---so that the
contribution of each signal to each query class is separable.

\textbf{Grounding} is the proportion of answers entailed by the records
retrieved for them, judged independently by two graders with agreement reported
as Cohen's $\kappa$.

\textbf{Abstention} is reported as a confusion matrix over all queries, crossing
whether a supporting memory exists with whether the system answered or declined.
The two error cells---answering when nothing supports the query, and declining
when a supporting memory was present---are reported separately, since they carry
different costs for a memory assistant.

\textbf{Response time} decomposes as
\begin{equation}
  T_{\text{total}} = T_{\text{stt}} + T_{\text{embed}} + T_{\text{retrieve}}
    + T_{\text{llm}} + T_{\text{tts}},
  \label{eq:latency}
\end{equation}
reported as median and 95th percentile per stage.

The corpus, the annotation schema, the difficulty audit and the harness
implementing these measures are released with the paper. \Cref{sec:results}
reports the survey, the transcription evaluation and the corpus construction;
the retrieval, grounding, abstention and latency measurements are specified and
implemented but not yet obtained, and we state this explicitly rather than
report partial figures.

%% file: tables/tab_corpus.tex
%% Gold corpus composition. Non-floating, self-contained.
\begingroup
\colorlet{growalt}{black!4}
\colorlet{growhead}{black!10}
\centering
\footnotesize
\setlength{\tabcolsep}{6pt}
\renewcommand{\arraystretch}{1.25}
\captionof{table}{Composition of the \sysname{} recall corpus: 25 conversations
  (503--642 words each), 160 gold memories and 250 queries. Mean and maximum
  content-word Jaccard overlap are measured between each query and the segment
  that supports it; lower values mean lexical matching alone is less able to
  resolve the query.}
\label{tab:corpus}

\begin{tabular}{@{}l r r r r@{}}
\toprule
\rowcolor{growhead}
\textbf{Query class} & \textbf{Count} & \textbf{Share} &
\textbf{Mean overlap} & \textbf{Max overlap} \\
\midrule
Exact identifier & 36 & 14.4\% & 0.203 & 0.500 \\
\rowcolor{growalt}
Paraphrase       & 36 & 14.4\% & 0.016 & 0.154 \\
Person-anchored  & 36 & 14.4\% & 0.072 & 0.250 \\
\rowcolor{growalt}
Temporal         & 36 & 14.4\% & 0.027 & 0.154 \\
Place-anchored   & 36 & 14.4\% & 0.116 & 0.438 \\
\rowcolor{growalt}
\textit{Unsupported} & 70 & 28.0\% & \textit{n/a} & \textit{n/a} \\
\midrule
\textbf{Total} & \textbf{250} & \textbf{100\%} & & \\
\bottomrule
\end{tabular}
\par\endgroup
\vspace{0.9\baselineskip}

%% file: sections/08_results.tex
\section{Results}
\label{sec:results}

\subsection{Survey: the Problem and the Binding Constraint}
\label{sec:results:survey}

\begin{figure}[htbp]
  \centering
  \includegraphics[width=\textwidth]{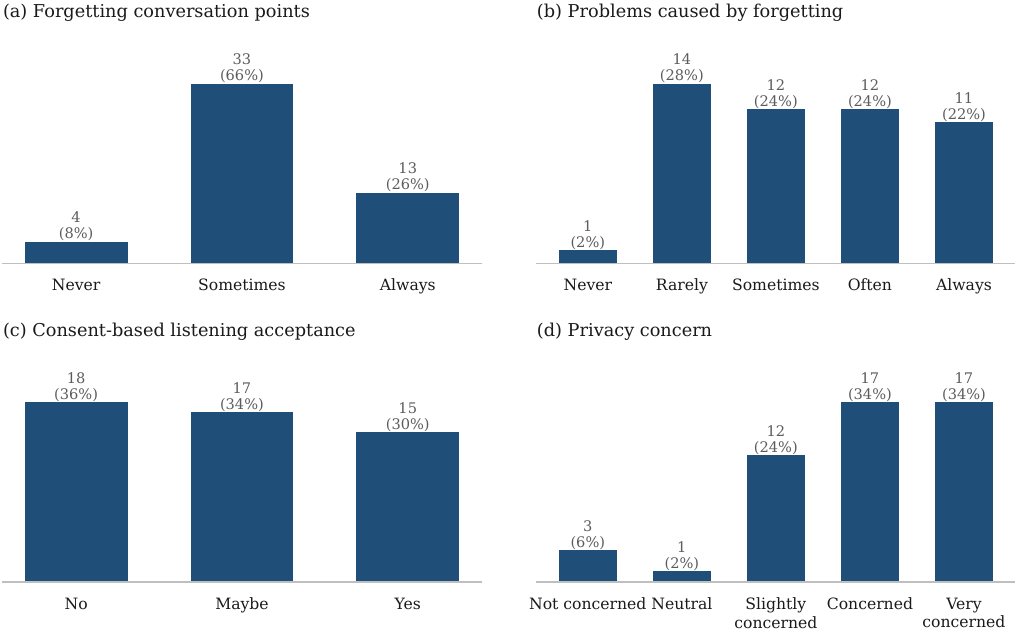}
  \caption{Survey responses ($n=50$): frequency of forgetting conversational
    detail, problems caused by it, acceptance of consent-based listening, and
    level of privacy concern.}
  \label{fig:surveypanels}
\end{figure}

\input{tables/tab_survey_summary}

Forty-six of fifty participants (92\%) reported forgetting important points from
conversations at least sometimes, and thirty-five (70\%) reported that this
causes them problems at least sometimes. Conversational forgetting is therefore
not confined to the clinical populations of \cref{sec:introduction}; it is a
general condition that becomes consequential at the margins.

The constraint on adoption is privacy rather than accuracy. Forty-six
participants (92\%) expressed at least slight concern about a system that stores
conversation history, and open responses returned repeatedly to data security
and unauthorised access rather than to whether the system would answer
correctly. Nonetheless thirty-two (64\%) answered ``yes'' or ``maybe'' to a
consent-based listening assistant, and thirty-eight (76\%) were neutral or
comfortable with a wearable interface---indicating that the concern is
conditional rather than prohibitive, and that it attaches to control over the
record rather than to capture as such. This is the finding that moved visible
recording state, reviewable records and reliable deletion from features to
design goals in \cref{sec:goals}.

\begin{figure}[htbp]
  \centering
  \includegraphics[width=0.92\textwidth]{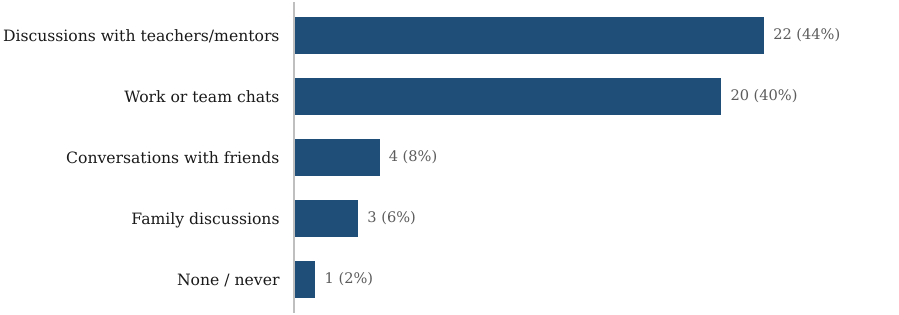}
  \caption{Situations in which participants expected \sysname{} to be most
    useful ($n=50$).}
  \label{fig:usecases}
\end{figure}

Expected usefulness concentrated on structured, information-dense settings:
discussions with teachers or mentors (22 participants, 44\%) and work or team
conversations (20, 40\%) together account for 42 of 50 responses. These are
contexts where spoken commitments are made and where failing to retain them has
a visible cost.

\subsection{Transcription Robustness}
\label{sec:results:asr}

\input{tables/tab_wer}

\begin{figure}[htbp]
  \centering
  \includegraphics[width=\textwidth]{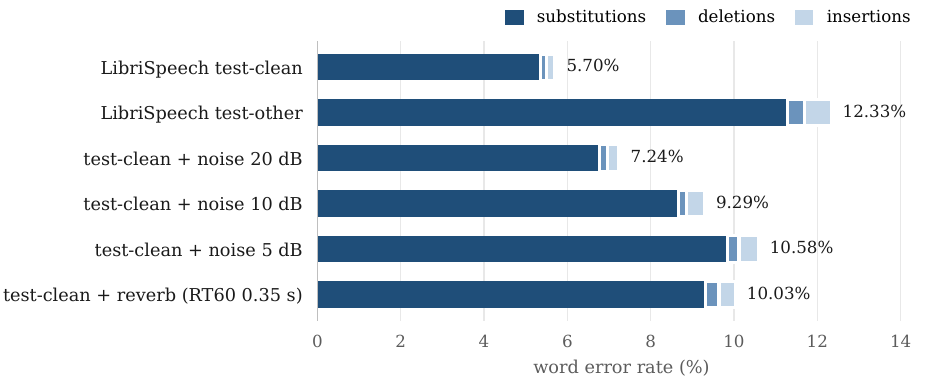}
  \caption{Composition of transcription error by acoustic condition. Each bar is
    segmented into substitutions, deletions and insertions, and labelled with
    total word error rate. Deletions remain below 0.4\% in every condition: the
    recogniser does not fall silent as conditions worsen, it substitutes. The
    code-switched condition is omitted here because at 77.36\% it is off-scale
    and, as \cref{tab:wer} notes, reflects a script mismatch rather than a
    comparable recognition result.}
  \label{fig:wer}
\end{figure}

\Cref{tab:wer} and \cref{fig:wer} report the capture stage across acoustic
conditions. Four
observations bear on the system's design.

\textbf{Speaker and channel variation costs more than it is usually credited
with.} Moving from \textsf{test-clean} to \textsf{test-other}---the same task,
read speech, with harder speakers and recording conditions---raises WER by a
factor of 2.2, from 5.70\% to 12.33\%, at identical model and decoding settings.
Any robustness claim quoted against clean read speech alone understates what a
deployed system encounters.

\textbf{Noise degrades recognition monotonically, and almost entirely through
substitution.} WER rises from 5.70\% unaugmented to 7.24\%, 9.29\% and 10.58\%
at 20, 10 and 5\,dB SNR, while the deletion rate stays between 0.15\% and 0.27\%
throughout. The model does not stop emitting words under noise; it emits the
wrong ones. This matters downstream, and it is the opposite of the intuitive
failure mode: a transcript degraded by noise reaches retrieval as \emph{wrong
index terms} rather than as missing content, so lexical matching fails silently
on a plausible-looking transcript rather than conspicuously on a sparse one.

\textbf{Mild reverberation is as damaging as substantial additive noise.} A
light synthetic reverberation (RT60 $=0.35$\,s) produces 10.03\% WER, close to
the 9.29\% measured at 10\,dB SNR, despite a far better signal-to-noise ratio.
Reverberant overlap is what two people speaking in an ordinary room actually
produce, which makes it the more realistic acoustic failure mode for a
voice-first assistant---and the more dangerous one, because the resulting
transcript remains fluent while its content is wrong.

\textbf{Code-switched speech exposes a script-handling problem distinct from
recognition difficulty.} On the Hindi--English test set the configuration
returns 77.36\% WER with a 9.76\% insertion rate, against 0.15--0.64\% for
English conditions. That error signature is characteristic of a script
mismatch: the reference transcripts are in Devanagari while the model emits
Latin script, so the normaliser compares codepoints from different writing
systems and much of the measured quantity is not recognition error. We report
the figure as an upper bound and do not treat it as a Hindi--English word error
rate. The substantive finding is the one the number reveals rather than
quantifies: a deployment in a code-switching setting cannot assume that
reference and hypothesis share a script, and must specify transliteration as
part of the transcription contract. Establishing a defensible error rate for
Roman-script Urdu--English speech requires either transliterated references or a
Devanagari-capable recogniser, and is left to future work.

Two limits apply to this study. Coverage is partial---8\,171 utterance-condition
pairs were transcribed of the 19\,175 in the design---and the per-condition
subsets are contiguous shards rather than random samples, so the ordering of
conditions and the composition of the error are the robust claims, not the third
decimal place. The two unaugmented rows are labelled clean in the sense of
unaugmented; \textsf{test-other} is not acoustically clean by construction, and
the two should not be pooled.

\subsection{What Building the Recall Corpus Revealed}
\label{sec:results:corpus}

The difficulty audit of \cref{sec:evalmethod:corpus} was intended as a quality
gate. It produced a methodological finding of its own, because the queries that
failed it were not the ones expected to.

Paraphrase queries were straightforward to make genuinely lexically distant from
their supporting segments: restating a proposition in different words is a
well-defined rewriting task, and the final corpus reaches a mean content-word
Jaccard overlap of 0.016 for that class (\cref{tab:corpus}). The class that
resisted was \emph{temporal}. Twenty-four of thirty-six temporal queries
initially had their supporting segment ranked inside the top three by embedding
similarity alone, meaning that a question whose only real anchor is a time was
nonetheless answerable by content similarity. The cause is structural: a short,
concrete segment tends to be the nearest neighbour of a generic question within
a short conversation, simply because nothing else in the conversation is close
to it either. Repairing this required adding decoy segments that restate the
query's generic scaffolding without containing the answer, over three rounds of
revision.

This generalises beyond our corpus. A retrieval benchmark that checks only
lexical overlap between paraphrase queries and their evidence will ship temporal
questions that content similarity alone resolves, and will consequently
understate how much work non-semantic signals are doing. We report the per-class
overlap distribution in \cref{tab:corpus} so that the corpus's difficulty is
inspectable rather than asserted.

A second observation concerns portability. Two temporal queries satisfied the
audit under one embedding model and failed it under another, and were revised
until they passed both. A difficulty audit is therefore relative to the encoder
it was run with, and a corpus validated for one encoder is not automatically
valid for another---a property that any released retrieval benchmark should
state alongside its data.

\subsection{Pilot Observations}
\label{sec:results:pilot}

\input{tables/tab_pilot_observations}

Participants understood the capture-then-ask workflow after a short
introduction. The interactions that worked most consistently were direct recall
requests: what a person said, which task was assigned, when an appointment
falls, where an object was placed. Short answers were markedly easier to use than long
ones, which supports G2 as a requirement rather than a preference. The difficulties observed were background noise, overlapping
speech, incomplete capture of context, and mismatch between the words a
participant used in a query and the words the stored record contained. The last
of these is the paraphrase case of \cref{tab:queryclasses} appearing in
practice, and the first two are the acoustic conditions quantified in
\cref{tab:wer}. Comfort was conditional on understanding what was being
recorded, consistent with the survey.

\subsection{Measurements Not Yet Obtained}
\label{sec:results:pending}

The retrieval, grounding, abstention and latency measures defined in
\cref{sec:evalmethod:measures} are implemented against the released corpus but
are not reported here, and we state that explicitly rather than offer partial
figures. Consequently this paper makes no empirical claim about retrieval
quality, answer faithfulness, abstention correctness or response time; the
contributions it does evidence are the problem characterisation, the
transcription robustness profile, the corpus and audit methodology, and the
system design itself. \Cref{sec:limitations} returns to what this leaves open.

%% file: tables/tab_survey_summary.tex
\begin{table}[t]
  \centering
  \small
  \caption{Survey response summary ($n=50$).}
  \label{tab:survey}
  \begin{tabularx}{\linewidth}{@{}X r r@{}}
    \toprule
    \textbf{Finding} & \textbf{Count} & \textbf{\%} \\
    \midrule
    Forget conversation points sometimes or always      & 46/50 & 92 \\
    Forgetting causes problems sometimes/often/always   & 35/50 & 70 \\
    Yes/Maybe to a consent-based listening assistant    & 32/50 & 64 \\
    At least slightly concerned about privacy           & 46/50 & 92 \\
    Neutral or comfortable with a wearable interface    & 38/50 & 76 \\
    \bottomrule
  \end{tabularx}
\end{table}

%% file: tables/tab_wer.tex
%% Transcription robustness. Non-floating, self-contained.
\makeatletter
\@ifundefined{NC@rewrite@Y}{\newcolumntype{Y}{>{\raggedright\arraybackslash}X}}{}
\makeatother
\begingroup
\colorlet{wrowalt}{black!4}
\colorlet{wrowhead}{black!10}
\colorlet{wrowflag}{remoraink!8}
\centering
\footnotesize
\setlength{\tabcolsep}{5pt}
\renewcommand{\arraystretch}{1.25}
\captionof{table}{Word error rate of the transcription stage on public corpora
  with published reference transcripts. $S$, $D$ and $I$ are substitution,
  deletion and insertion rates. $n$ is the number of utterances transcribed, of
  the split size in parentheses. The tinted row is reported as an upper bound
  only: its references are in Devanagari while the model emits Latin script, so
  a large share of the measured error is script mismatch rather than
  recognition error (\cref{sec:results:asr}). Conditions are reported
  separately and are not pooled.}
\label{tab:wer}

\begin{tabularx}{\textwidth}{@{}>{\raggedright\arraybackslash}p{1.7cm} >{\raggedright\arraybackslash}p{2.6cm} >{\raggedright\arraybackslash}p{2.1cm} >{\raggedright\arraybackslash}p{1.5cm} r r r r@{}}
\toprule
\rowcolor{wrowhead}
\textbf{Condition} & \textbf{Corpus} & \textbf{Augmentation} & \textbf{$n$} &
\textbf{WER} & \textbf{$S$} & \textbf{$D$} & \textbf{$I$} \\
\midrule
read, clean  & LibriSpeech \textsf{test-clean} & none & 1\,249 (2\,620)
  & \textbf{5.70\%} & 5.36 & 0.15 & 0.18 \\
\rowcolor{wrowalt}
read, harder & LibriSpeech \textsf{test-other} & none & 1\,439 (2\,939)
  & \textbf{12.33\%} & 11.29 & 0.40 & 0.64 \\
noisy & \textsf{test-clean} & noise, 20\,dB SNR & 1\,319 (2\,620)
  & \textbf{7.24\%} & 6.77 & 0.20 & 0.26 \\
\rowcolor{wrowalt}
noisy & \textsf{test-clean} & noise, 10\,dB SNR & 1\,288 (2\,620)
  & \textbf{9.29\%} & 8.67 & 0.20 & 0.42 \\
noisy & \textsf{test-clean} & noise, 5\,dB SNR & 672 (2\,620)
  & \textbf{10.58\%} & 9.85 & 0.27 & 0.46 \\
\rowcolor{wrowalt}
reverberant & \textsf{test-clean} & reverb, RT60 0.35\,s & 920 (2\,620)
  & \textbf{10.03\%} & 9.31 & 0.33 & 0.40 \\
\rowcolor{wrowflag}
code-switched & MUCS 2021 Hindi--English & none & 1\,284 (3\,136)
  & \textbf{77.36\%}$^{\dagger}$ & 61.11 & 6.50 & 9.76 \\
\bottomrule
\end{tabularx}
\par\endgroup
\vspace{0.3\baselineskip}
{\footnotesize $^{\dagger}$Script-mismatch upper bound; see
\cref{sec:results:asr}.}
\vspace{0.7\baselineskip}

%% file: tables/tab_pilot_observations.tex
\begin{table}[t]
  \centering
  \small
  \caption{Pilot testing observations with 10 memory-loss participants.}
  \label{tab:pilot}
  \begin{tabularx}{\linewidth}{@{}l X@{}}
    \toprule
    \textbf{Aspect} & \textbf{Observation} \\
    \midrule
    Audio capture      & Worked in normal indoor settings; noise reduced quality \\
    Transcription      & Usable transcripts for clear speech \\
    Memory extraction  & Useful tasks, names, appointments and summaries extracted \\
    Retrieval          & Direct questions usually retrieved relevant memories \\
    Response quality   & Short answers were easier to understand \\
    User comfort       & A short explanation was required before use \\
    Main concerns      & Privacy, accidental recording, unauthorised access \\
    \bottomrule
  \end{tabularx}
\end{table}

%% file: sections/09_discussion.tex
\section{Discussion}
\label{sec:discussion}

\subsection{Privacy Is the Binding Constraint, and It Is Conditional}
\label{sec:discussion:privacy}

The survey's most consequential result is not that privacy concern is high but
that it coexists with willingness. Ninety-two per cent of participants
expressed concern about a system retaining conversation history, and
sixty-four per cent would nonetheless accept a consent-based listening
assistant. A system treating those as contradictory would conclude that the
product is unwanted; a system treating privacy as a design surface concludes
something more useful, which is that acceptance is purchased with control.

This has an architectural consequence that \cref{sec:goals} drew and
\cref{sec:design} implemented: deletion must be real. If the user's assurance
is control over the record, then a delete that leaves an embedding in a vector
index has not merely failed to deliver a feature---it has invalidated the basis
on which the system was accepted. The consolidation onto a single transactional
store follows from the survey finding, not from a performance argument, and we
regard that as the cleanest example in this work of an empirical result
determining an engineering decision.

\subsection{Transcription Failure Is Substitution, Not Omission}
\label{sec:discussion:asr}

\Cref{sec:results:asr} found that degradation under noise is almost entirely
substitutive: deletions remain below 0.27\% across a 15\,dB range of SNR while
WER nearly doubles. For a memory assistant this is worse than the alternative
would be. A transcript with gaps advertises its own unreliability, and both a
lexical index and a reader can treat the gap as missing information. A
transcript with substitutions is fluent, confident and wrong: it enters the
store as a well-formed memory record, is embedded as though it meant what it
says, and is retrieved later as evidence.

Grounding does not protect against this. The citation contract of
\cref{sec:retrieval:generation} guarantees that an answer is supported by a
stored record; it cannot guarantee that the record faithfully represents what
was said. The protection that does apply is the one the interface already
provides---retaining the verbatim segment beside the extracted summary
(\cref{app:interface}), so a user who doubts an answer can read the source.
Provenance, on this reading, is not only an auditing mechanism against model
fabrication but the only available defence against upstream transcription
error.

\subsection{Reverberation, Not Noise, Is the Realistic Acoustic Threat}
\label{sec:discussion:reverb}

A light reverberation produced error comparable to 10\,dB additive noise
despite a far better signal-to-noise ratio. Benchmarks commonly characterise
robustness through additive noise, but the acoustic condition a conversational
memory assistant actually meets is two people speaking in an ordinary room,
which is a reverberation and overlap problem. This suggests that robustness
claims for systems of this kind should be quoted against reverberant conditions
rather than SNR sweeps, and it is a measurement that costs very little to add.

\subsection{Why Hybrid Retrieval Is a Coverage Requirement}
\label{sec:discussion:hybrid}

The query taxonomy of \cref{tab:queryclasses} and the overlap distribution of
\cref{tab:corpus} together make an argument that does not depend on the
retrieval measurements still outstanding. Two of six query classes---temporal
and, in large part, person-anchored---carry almost no topical content: their
mean content-word overlap with the supporting segment is 0.027 and 0.072
respectively. A retriever ranking only on content similarity has nothing to
rank them by. Conversely, exact-identifier queries turn on rare tokens that
embeddings place poorly, with the highest lexical overlap of any class (0.203).
Each family of signals is therefore necessary for a class the other cannot
serve, which is why we characterise hybrid retrieval as a coverage requirement
rather than a quality improvement. Quantifying the margin is the purpose of the
ablation specified in \cref{sec:evalmethod:measures}.

\subsection{Relationship to Memory Assistants and Agent Memory}
\label{sec:discussion:related}

\sysname{} adopts its interaction commitments from the wearable memory
assistants of \cref{sec:related:llm}---short queries, concise answers, a
confidence-gated proactive mode---and departs from them in two respects. It
removes the hardware precondition, which those systems share and which excludes
the populations with the strongest need. And it adds a commitment about the
evidentiary status of the answer: that it is returned with the material it was
drawn from, and withheld when that material is absent.

Relative to the agent memory architectures of \cref{sec:related:agents}, the
difference is what the store is for. An agent's memory serves the agent and
need never be inspected; a memory assistant's store is a record of the user's
life, which they must be able to read, correct and delete. That requirement is
what makes a single transactional system of record a design constraint here
rather than an implementation preference, and it is the reason the rank-fusion
formulation of \cref{eq:hybrid} can be evaluated in one query plan rather than
assembled from the outputs of independent services.

%% file: sections/10_threats_to_validity.tex
\section{Threats to Validity}
\label{sec:threats}

\paragraph{Construct validity.} The survey measures stated intention, not
behaviour. Expressed willingness to accept consent-based listening is a poor
predictor of sustained use once recording is continuous and ambient, and the
76\% who were comfortable with a wearable interface were responding to a
description rather than to a device. The pilot partially addresses this by
exposing participants to a working system, but over minutes rather than weeks.
Separately, the transcription study measures a component on read and
conversational benchmark speech, which is a proxy for---not an instance
of---the audio \sysname{} will encounter in homes and clinics.

\paragraph{Internal validity.} Pilot sessions were conducted in controlled
settings with the research team present, and scenarios were scripted. Both
raise demand characteristics: participants knew which system was being
evaluated and what it was supposed to do. The transcription study is free of
this, but introduces its own: condition subsets were taken as contiguous shards
of each split rather than as random samples, so between-condition comparisons
rest on the assumption that the shard is representative of the split. We regard
the ordering of conditions and the composition of the error as robust to this,
and precise rates as not.

\paragraph{External validity.} Fifty survey respondents and ten pilot
participants, recruited in one city and largely from one institutional network,
do not support generalisation to other populations, age distributions or
healthcare contexts. The speech corpora are English read speech and
Hindi--English conversational speech; neither is Roman-script Urdu--English as
spoken in Karachi, which is the deployment setting the paper describes. The
code-switched figure should therefore be read as evidence that script handling
is a problem class, not as a performance estimate for the target population.

\paragraph{Conclusion validity.} With $n=10$ the pilot supports description
only, and we report no inferential statistics from it. The survey proportions
are reported without confidence intervals; at $n=50$ the 95\% interval on a
92\% proportion is approximately $\pm 7$ percentage points, which does not
affect the qualitative conclusions but should temper reading of the exact
values. Most importantly, the retrieval, grounding, abstention and latency
measures are not reported at all (\cref{sec:results:pending}), so no conclusion
in this paper rests on them, and none should be inferred.

%% file: sections/11_limitations_future_work.tex
\section{Limitations and Future Work}
\label{sec:limitations}

The clearest limitation of this paper is the gap between what \sysname{}
specifies and what has been measured. The retrieval, grounding, abstention and
latency measures of \cref{sec:evalmethod:measures} are implemented against a
released corpus but unreported, so the central claim---that hybrid retrieval
with grounded generation produces more trustworthy recall than single-signal
search---remains specified rather than demonstrated. The corpus, the annotation
schema, the difficulty audit and the measurement harness are released precisely
so that this gap is closeable, by us or by others.

Three limitations concern the system rather than the evaluation. First,
abstention is enforced through the generation contract over retrieved evidence
rather than by a retrieval-side score threshold. Adding a minimum-score
criterion would make the decision to decline independent of the generator
altogether, which is a stronger guarantee than the one currently implemented,
and is the single change we would prioritise. Second, provenance is recorded at
the granularity of the capture and its transcript, not of the individual
segment that produced each memory: an answer can be traced to the conversation
it came from, but per-memory segment attribution requires the extraction stage
to assign sources explicitly, which it does not yet do. Finer attribution would
both sharpen the user-facing citation and tighten any provenance-based
retrieval metric. Third, the deletion guarantee holds within the system of
record but has not been audited end to end against derived artefacts outside it,
such as provider-side logs.

Three concern the setting. The transcription study establishes that script
mismatch, not recognition difficulty alone, dominates measured error on
code-switched references; a defensible error rate for Roman-script
Urdu--English speech requires either transliterated references or a
script-aware recogniser, and that measurement is outstanding. Reverberation
proved as damaging as substantial additive noise, which suggests that
dereverberation and overlap handling deserve more attention than noise
robustness in this application. And the privacy properties that the survey
identifies as decisive have been designed for but not tested with bystanders,
who are the parties least represented in any of our studies and most exposed by
ambient capture.

Future work follows these directly: completing the measurement programme on the
released corpus; implementing segment-level attribution and a retrieval-side
abstention threshold; establishing a transliteration-aware transcription
contract for code-switched speech; and a longitudinal deployment in which
acceptance is observed over weeks rather than elicited in a questionnaire,
with caregiver involvement where appropriate and with bystander consent treated
as a first-class design problem rather than a disclosure.

%% file: sections/12_conclusion.tex
\section{Conclusion}
\label{sec:conclusion}

This paper presented \sysname{}, a voice-first memory assistant that converts
consented speech into structured, auditable memory records and answers
questions from them. Its design rests on three commitments: that recall should
cost the user less attention than the lapse it repairs; that an answer should
either carry the evidence it came from or be withheld; and that the record of a
person's conversations must remain inspectable, correctable and genuinely
erasable by them.

The evidence reported here supports the problem and characterises the capture
stage. Conversational forgetting is common well beyond clinical populations,
and the constraint on adopting a system that addresses it is privacy rather
than accuracy---a concern that proved conditional rather than prohibitive, and
that attaches to control over the stored record. Transcription degrades under
realistic acoustics almost entirely through substitution rather than omission,
which means capture error reaches a memory store as confident, well-formed and
wrong; and mild reverberation proved as damaging as substantial additive noise,
making conversational overlap rather than background noise the acoustic
condition such systems should be evaluated against. Building the evaluation
corpus produced a finding of its own: temporal queries, not paraphrases, are
the class that retrieval benchmarks most easily make accidentally easy.

The retrieval, grounding and abstention measurements that would test the
system's central claim are specified and implemented but not yet obtained, and
we have reported that rather than substitute partial figures. The corpus,
schema, audit and harness are released so the claim can be settled.

The broader argument is that a memory assistant is held to a different standard
than a general one. A fluent invention from a chatbot is an error the user can
check against the world; a fluent invention about the user's own past is a
corrupted memory, checkable only against the faculty the system exists to
compensate for. That asymmetry is why provenance and abstention belong in the
architecture rather than in the prompt, and why a system of this kind should be
judged not only by what it can recall, but by how reliably it declines.

%% file: sections/13_backmatter.tex
\section*{Data and Materials Availability}
\label{sec:availability}

The \sysname{} source, the recall corpus of \cref{sec:evalmethod:corpus} with
its annotation schema, the corpus difficulty audit, the evaluation harness
implementing the measures of \cref{sec:evalmethod:measures}, the survey
instrument and pilot questionnaire, the anonymised survey responses, and the
prompt templates of \cref{app:prompts} are available in the project
repository~\citep{karim2026remorarepo}. Raw per-utterance transcription outputs
and edit counts for \cref{tab:wer} are included, so every reported rate can be
recomputed. The speech corpora used in \cref{sec:results:asr} are third-party
resources and are not redistributed; the identifiers, splits and licences are
given in \cref{sec:evalmethod:asr}.

\section*{Ethics Statement}
\label{sec:ethics}

Both studies were conducted with informed consent. Survey participation was
voluntary, anonymous and unremunerated, and respondents were told how their
responses would be used before beginning. Pilot participants were recruited on
the basis of self-reported memory difficulty rather than clinical diagnosis; no
medical records were consulted and no diagnostic information was collected. The
procedure, the right to stop at any point and the handling of recorded audio
were explained before consent was taken, and participants were accompanied by a
family member or caregiver where they preferred. Scenario conversations were
scripted and performed for the study; no third party was recorded without
knowledge, and no personal or clinical content from participants' own lives was
captured. Audio and transcripts were held only for the duration of the session
and deleted afterwards, and no identifying material appears in the released
data. \sysname{} is an assistive prototype and was presented as such: no
diagnostic, therapeutic or clinical benefit was claimed to participants at any
point.

\section*{Acknowledgments}
\label{sec:acks}

We thank the Department of Computer Science at Salim Habib University for
supporting this project, the survey respondents for their time, and the pilot
participants and their families for their patience with an early prototype.

%% file: appendix/A_survey_instrument.tex
\section{Survey Instrument and Response Options}
\label{app:survey}

The survey was administered to fifty participants and is reproduced verbatim
below. Items 1--2 collected demographic context, items 3--5 established the
frequency and cost of conversational forgetting, items 6--7 and 10 elicited
privacy and acceptance constraints, and items 8--9 identified where
participants expected a memory assistant to be useful. Responses were
anonymous.

\begin{enumerate}[leftmargin=1.6em,itemsep=0.3em]
  \item What is your age group?
  \item What is your gender?
  \item How often do you forget important points from conversations with friends,
    family or teachers? \emph{(Never / Sometimes / Always)}
  \item How do you usually try to remember important conversations?
    \emph{(Mental notes / Writing down / Messaging yourself / Recording audio /
    Other)}
  \item When you forget something from a past conversation, how often does it
    cause you problems? \emph{(Never / Rarely / Sometimes / Often / Always)}
  \item Would you feel comfortable having a wearable or app listen to your
    conversations, with consent, to help you remember them later?
    \emph{(Yes / Maybe / No)}
  \item How concerned are you about privacy when using such a memory-assisting AI?
    \emph{(Not concerned / Neutral / Slightly concerned / Concerned /
    Very concerned)}
  \item In what situations do you think a memory assistant like \sysname would
    help you most?
  \item How comfortable are you with a wearable device, such as bone-conduction
    headphones or a discreet microphone, compared with a phone-only app?
  \item What are your main concerns, if any, about using an AI assistant that
    remembers your conversations for you?
\end{enumerate}

%% file: appendix/B_pilot_questionnaire.tex
\section{Pilot Evaluation Questionnaire}
\label{app:questionnaire}

Pilot participants completed the following items after their session, rated on
a 7-point Likert scale (1 = strongly disagree, 7 = strongly agree) except where
a choice is offered. The items map onto the design goals of
\cref{sec:goals:list}: items 1 and 6 to G2 (minimal output), item 4 to G1
(minimal input), items 2 and 3 to recall usefulness, items 5 and 8 to G4 (user
control and privacy), and item 9 to G5 (practical deployment).

\begin{enumerate}[leftmargin=1.6em,itemsep=0.25em]
  \item The response from \sysname was short enough to understand quickly.
  \item The answer from \sysname helped me remember the missing information.
  \item I felt more confident in recalling the conversation after using \sysname.
  \item I found it easy to ask \sysname a question.
  \item I understood when the system was recording or processing information.
  \item The system did not feel too distracting during the task.
  \item I would find \sysname useful in daily life.
  \item I felt comfortable with \sysname storing my conversation memory.
  \item I would prefer a phone-only interface, a wearable interface, or both.
  \item My main concern about \sysname is: privacy, accuracy, comfort, cost, or
    other.
\end{enumerate}

%% file: appendix/C_pilot_scenarios.tex
\section{Pilot Test Scenarios}
\label{app:scenarios}

Each pilot session worked through the scenario types below. The first five
have a correct answer recoverable from the captured conversation; the sixth has
none, and the correct behaviour is an explicit statement that the memory is
unavailable. The sixth was included in every session, so that abstention was
observed rather than assumed.

\begin{description}[leftmargin=0pt,style=unboxed,itemsep=0.3em]
  \item[\textbf{Conversation recall.}] The participant records a short
    conversation and later asks what was discussed.
  \item[\textbf{Task recall.}] A task or deadline is mentioned; the participant
    asks what they need to do.
  \item[\textbf{Appointment recall.}] A date, time or visit instruction is
    mentioned and requested later.
  \item[\textbf{Name recall.}] A person's name or role is mentioned; the
    participant asks who the person was.
  \item[\textbf{Object-location recall.}] A spoken statement mentions where an
    object was kept; the participant asks for the location.
  \item[\textbf{Unsupported query.}] The participant asks about information that
    was never captured, testing whether \sysname refuses rather than guesses.
\end{description}

%% file: appendix/D_prompt_templates.tex
\section{Prompt Templates}
\label{app:prompts}

The prompts below are the control logic of the grounding contract described in
\cref{sec:retrieval:generation}, and are reproduced verbatim as used. They are
research-prototype prompts rather than a fixed production interface, and are
given so that the behaviour reported in this paper can be reproduced or
contested.

\promptbox{Grounded recall prompt}{%
You are \sysname, a personal memory recall assistant. Answer the user's question
only using the retrieved memory context. If the answer is not present in the
context, say that the memory is not available. Keep the answer short, clear, and
suitable for spoken output. Do not invent personal details.}

\promptbox{Concise answer prompt}{%
Rewrite the answer so it contains only the words needed to satisfy the user's
memory need. Remove repeated context, unnecessary explanation, and details that
are not directly asked for. Preserve accuracy.}

\promptbox{Assisted / queryless recall prompt}{%
Given the recent conversation context, infer whether the user may need a memory
cue from stored memories. If the need is clear and confidence is high, formulate a
short retrieval query. If the need is unclear, return no suggestion.}

%% file: appendix/E_reproducibility.tex
\section{Reproducibility Details}
\label{app:repro}

\paragraph{Retrieval.} Candidate lists are produced by PostgreSQL full-text
search, \texttt{pgvector} cosine similarity and, where a memory carries a
location, a PostGIS proximity filter. The lists are combined by reciprocal rank
fusion (\cref{eq:hybrid}) with $\kappa = 60$, the conventional value, which was
not tuned. No per-signal weights are applied and no minimum-score cutoff is
imposed; the consequences of the latter are discussed in
\cref{sec:limitations}.

\paragraph{Generation.} Answers are produced from the retrieved records under
the prompts in \cref{app:prompts}. For the evaluation reported here, decoding
was pinned to temperature 0 with a fixed seed so that runs are reproducible;
the deployed configuration does not pin these, which is recorded here because
it is the only behavioural difference between the evaluated and deployed call.

\paragraph{Transcription study.} LibriSpeech \textsf{test-clean} and
\textsf{test-other} (OpenSLR SLR12, CC~BY~4.0); MUCS 2021 sub-task 2
Hindi--English test set (OpenSLR SLR104, CC~BY-SA~4.0); noise from the MUSAN
noise category (OpenSLR SLR17, CC~BY~4.0). Augmentation was deterministic at
16\,kHz: additive noise scaled so that full-clip noise RMS meets speech RMS at
20, 10 and 5\,dB SNR, and a synthetic reverberation with an exponentially
decaying impulse response, RT60 $=0.35$\,s, wet mix 0.6. Word error rate is
computed by word-level Levenshtein alignment after lowercasing, stripping
punctuation, expanding numerals and removing \texttt{<unk>} and \texttt{<noise>}
tokens.

\paragraph{Recall corpus.} Twenty-five conversations of 503--642 words, 160 gold
memories and 250 queries, in clean (12), code-switched (7) and noisy (6)
conditions, across clinic (5), kitchen (4), meeting (4), lecture (4), shop (4)
and phone (4) settings. The difficulty audit reported in
\cref{sec:results:corpus} was run with the same embedding model used by the
retrieval path; as \cref{sec:results:corpus} notes, such an audit is relative to
the encoder it was run with, and the corpus records which one.

\paragraph{Store and teardown.} Evaluation data is written to a dedicated
evaluation account so that it never mixes with user data, and is removed by a
single cascading delete. Ingestion is idempotent through content-derived
identifiers, so re-running the harness cannot duplicate memories.

%% file: appendix/F_interface.tex
\section{Interface Walkthrough}
\label{app:interface}

This appendix shows the \sysname{} mobile client referenced in
\cref{sec:impl:ui}. The screens are ordered along the path a memory takes:
capture, summarisation, recall, and inspection of what was stored.

\begin{figure}[htbp]
  \centering
  \includegraphics[width=\textwidth]{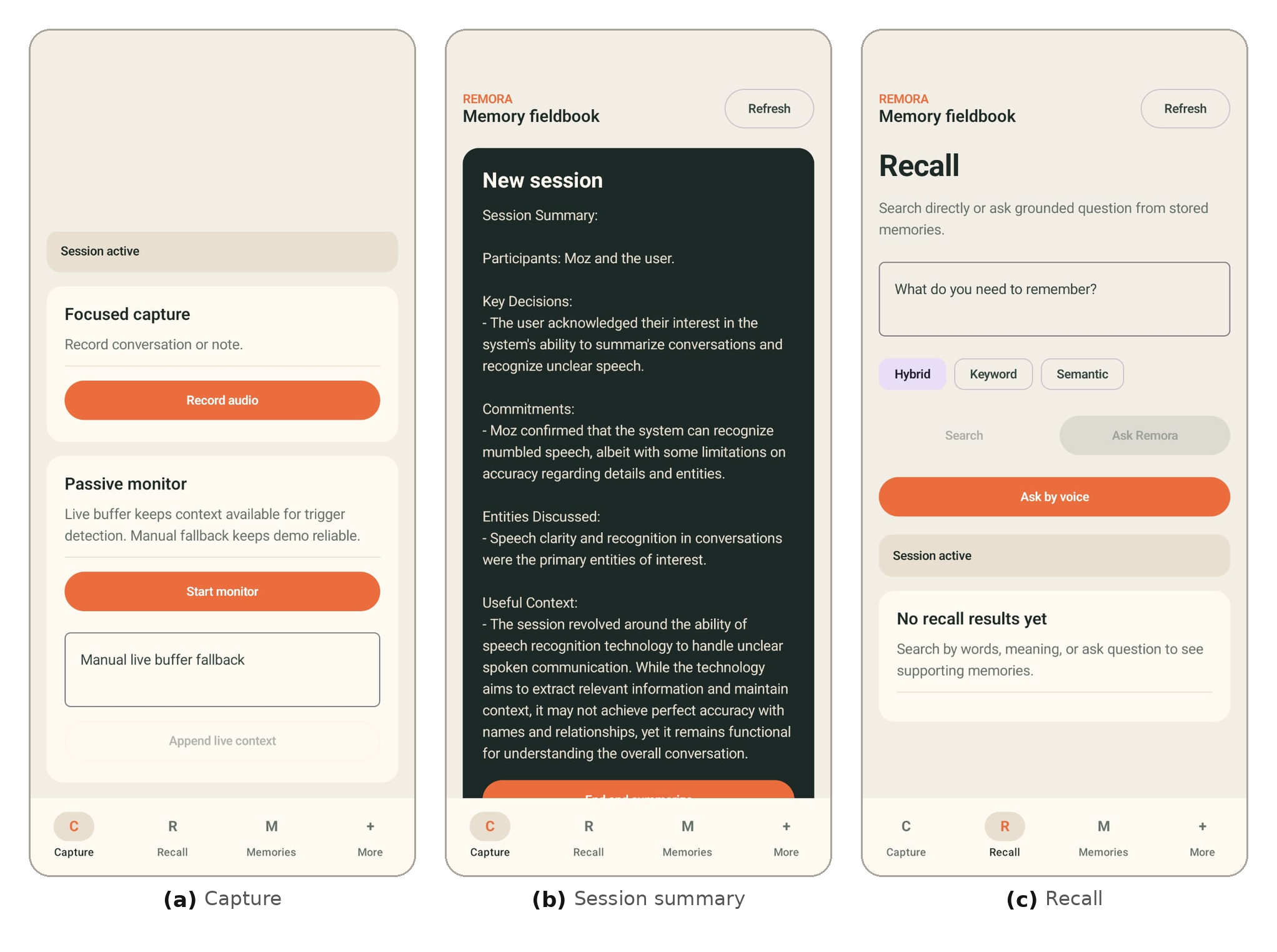}
  \caption{Capture and recall. \textbf{(a)}~Capture offers focused recording of
    a conversation and a passive monitor that keeps a rolling buffer for trigger
    detection; the \textsf{Session active} banner is persistent, which is the
    visible-state half of G4. \textbf{(b)}~When a session ends, \sysname{}
    produces a session summary organised by decisions, commitments and entities,
    giving the user a reviewable account of what was retained.
    \textbf{(c)}~Recall accepts a typed or spoken question and exposes the
    retrieval mode explicitly---\textsf{Hybrid}, \textsf{Keyword} or
    \textsf{Semantic}---so that the signals combined in \cref{eq:hybrid} can
    also be exercised in isolation.}
  \label{fig:ui-capture}
\end{figure}

\begin{figure}[htbp]
  \centering
  \includegraphics[width=\textwidth]{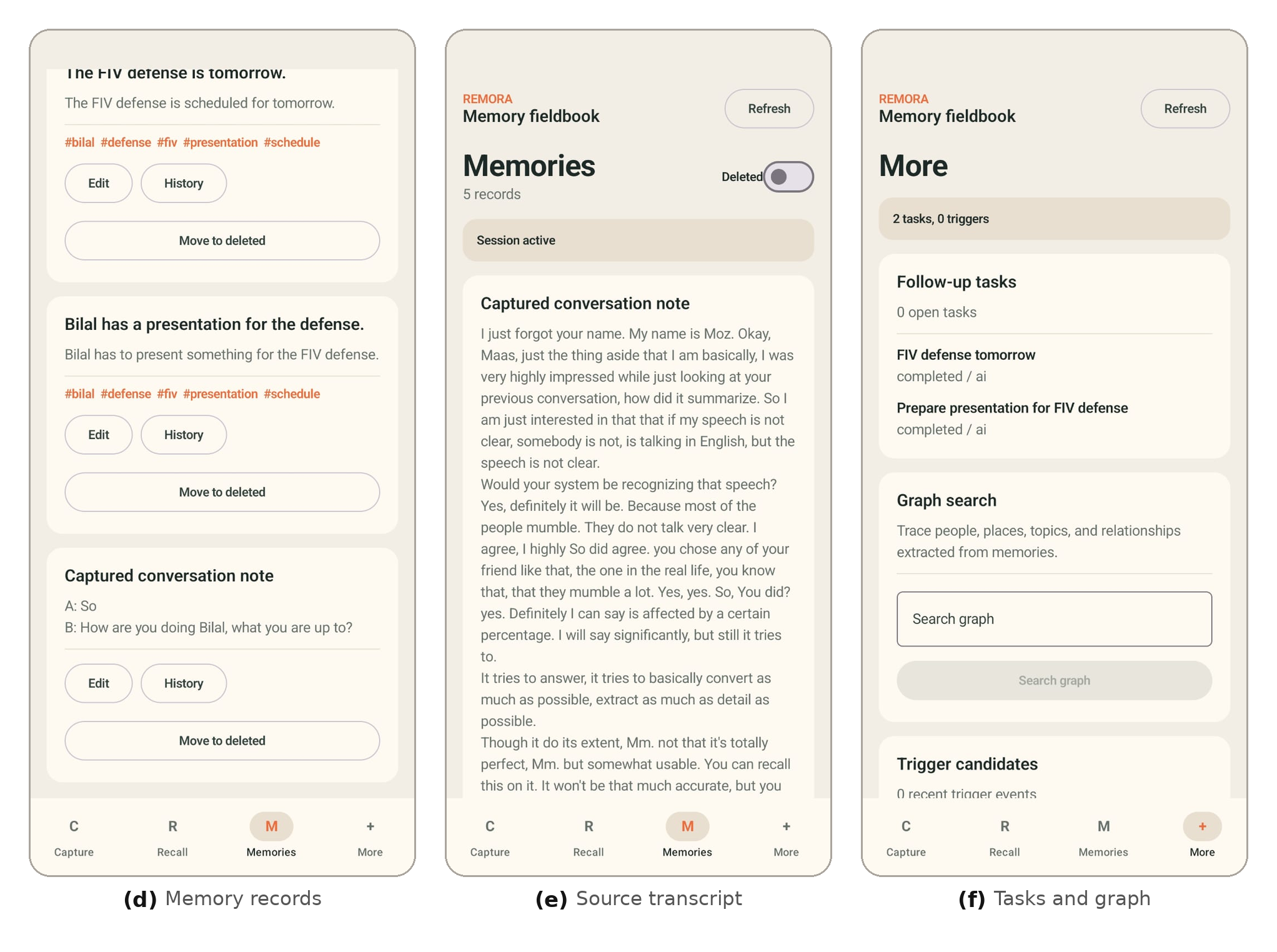}
  \caption{Stored memory and its provenance. \textbf{(d)}~Extracted records are
    listed with their tags and with per-record controls for editing, viewing
    revision history, and removal, satisfying the review-and-correct requirement
    of G4. \textbf{(e)}~The verbatim transcript is retained alongside the
    extracted records rather than discarded, which is what allows an answer to
    be traced to the source segment $x_i$ of \cref{eq:record} rather than only
    to a summary. \textbf{(f)}~Entities and relationships extracted from
    memories are searchable as a graph, which is the relational signal $C$ of
    \cref{eq:hybrid}; trigger candidates are the inferred memory needs that
    assisted recall evaluates against the gates of
    \cref{sec:retrieval:modes}.}
  \label{fig:ui-memory}
\end{figure}